\documentclass{aa}  

\usepackage[utf8]{inputenc} 
\usepackage[T1]{fontenc}
\usepackage{gensymb}
\usepackage{graphicx}
\usepackage{subfig}
\usepackage{txfonts}
\usepackage{colortbl}
\definecolor{pale}{RGB}{255,200,200}
\usepackage{hyperref}

\usepackage{color}

\begin{document} 

   \title{On the origin of the wide-orbit circumbinary giant planet HD 106906}
   
   \subtitle{A  dynamical scenario and its impact on the disk}

   \author{L. Rodet\inst{1} \and H. Beust\inst{1} \and M. Bonnefoy\inst{1}, A.-M. Lagrange\inst{1}, P. A. B. Galli\inst{1,2}, C. Ducourant\inst{3} \and R. Teixeira\inst{2}}
   \institute{Univ. Grenoble Alpes, CNRS, IPAG, F-38000 Grenoble, France\\
                \email{laetitia.rodet@univ-grenoble-alpes.fr}
                        \and
                Instituto de Astronomia, Geofísica e Ciências Atmosféricas, Universidade de São Paulo, Rua do Matão, 1226, Cidade Universitária, 05508-900, São Paulo - SP, Brazil
                \and
                Laboratoire d’Astrophysique de Bordeaux, Univ. Bordeaux, CNRS, B18N, allée Geoffroy Saint-Hilaire, 33615 Pessac, France}

   \date{Received 16 December 2016 / Accepted 4 March 2017}

 
  \abstract
   {A giant planet has been recently resolved at a projected distance of 730 au from the tight pair of young ($\sim$ 13 Myr) intermediate-mass stars HD~106906AB in the Lower Centaurus Crux (LCC) group. The stars are surrounded by a debris disk which displays a ring-like morphology and strong asymmetries at multiple scales.}
  {We aim to study the likelihood of a scenario where the planet formed closer to the stars in the disk, underwent inward disk-induced migration, and got scattered away by the binary star before being stabilized by a close encounter (fly-by).}
   {We performed semi-analytical calculations and numerical simulations (Swift\_HJS package) to model the interactions between the planet and the two stars. We accounted for the migration as a simple force. We studied the LCC kinematics to set constraints on the local density of stars, and therefore on the fly-by likelihood. We performed N-body simulations to determine the effects of the planet trajectories (ejection and secular effects) onto the disk morphology.}
   {The combination of the migration and mean-motion resonances with the binary star (often 1:6) can eject the planet. Nonetheless, we estimate that the fly-by hypothesis decreases the scenario probability to less than $10^{-7}$ for a derived local density of stars of 0.11 stars/pc$^{3}$. We show that the concomitant effect of the planet and stars trajectories induce spiral-features in the disk which may correspond to the observed asymmetries. Moreover, the present disk shape suggests that the planet is on an eccentric orbit.}
   {The scenario we explored is a natural hypothesis if the planet formed within a disk. Conversely, its low probability of occurrence and the fact that HD~106906~b shares some characteristics with other systems in Sco-Cen (e.g., HIP~78530, in terms of mass ratio and separation) may indicate an alternative formation pathway for those objects.}


   \maketitle

\section{Introduction}

More than 3500 exoplanets have been found in the last three decades\footnote{http://exoplanet.eu}, but few among them have been detected to be hundreds of astronomical units (au) from their star. As the development of direct imaging reveals more of those wide planetary-mass companions, classical theories of planet formation fail at explaining their origin. In the two scenarios, core accretion \citep{Accretion} and gravitational instability \citep{GI}, the planets form within the primordial gas disk. However, the limited extent of the disk (see eg Fig. 5 in Lieman-Sifry et al. \citeyear{DisquesScoCen}) does not enable the formation of a giant planet far away from its star. Thus, when the star around which orbits the very wide and massive HD~106906AB~b turned out to be a binary star \citep{Binaire}, it has been suggested that dynamical interactions could account for the current position of the planet (Lagrange et al. \citeyear{Binaire}, Wu et al. \citeyear{Wu}).

The planet HD 106906 (or also HIP 59960) is located at a distance of $103\pm4$ pc \citep{Hipparcos} and belongs to the Lower Centaurus Crux (LCC) group, which is a subgroup of the Scorpius–Centaurus (Sco-Cen) OB association \citep{HipparcosOB}. The LCC group has a mean age of 17~Myr, with an age-spread of about 10~Myr \citep{LCC}. In recent years, high contrast imaging has revealed the circumstellar environment of HD~106906AB: an $11\pm 2~M_J$ planet located at $732 \pm 30$~au in projected separation (Bailey et al. 2013) and an asymmetric debris disk nearly viewed edge-on, imaged by SPHERE \citep{DisqueSphere}, GPI and HST \citep{DisqueGPI} and MagAO \citep{Wu}. More recently, the binary nature of HD 106906 was inferred thanks to observations with the instruments HARPS and PIONIER \citep{Binaire}. It turns out to be a $13 \pm 2$~Myr old SB2 binary consisting of two F5 V-type stars with very similar masses. Table \ref{table:observations} summarizes the key characteristics of the system components. No further information is known about the orbit of the planet, which must have an orbital period of at least 3000 years. The binary orbit is also not much constrained yet, but given its short orbital period ($<100$ days), it will presumably be better known in the near future. 

\begin{table}[h]
        {\tiny
        \caption{Key characteristics of the HD 106906 system.}
        \label{table:observations}                            
        \begin{center}
        \begin{tabular}{c c c}        
        \hline\hline                 
        System component & Mass & Projected separation \\     
        \hline\\[-0.8em]                      
    HD 106906 AB &  $\sim 1.34$ and $\sim 1.37~ M_\odot$ $^a$ & 0.36-0.58 au $^a$\\
        HD 106906 b & $11 \pm 2~ M_J$ $^b$ & $732 \pm 30$ au $^b$\\
        Disk & $0.067 ~ M_\text{Moon}$ $^c$ & from $65 \pm 3$ to $\sim 550$ au $^d$ $^e$\\
        \hline                                   
        \end{tabular}
        \end{center}}\vspace{0.5cm}
        \small{References: $a$ -- Lagrange et al. \citeyear{Binaire}, $b$ -- Bailey et al. \citeyear{Decouverte}, $c$ -- Chen et al. \citeyear{ScoCen}, $d$ -- Kalas et al. \citeyear{DisqueGPI}, $e$ -- Lagrange et al. \citeyear{DisqueSphere}.}
\end{table}

The edge-on debris disk has an unusual shape: its luminous intensity has a very asymmetric profile. The longest peak, pointing westward, extends up to 550 au, while the east edge reaches 120 au only (see Figs. 1 and 3 of \citeauthor{DisqueGPI} \citeyear{DisqueGPI}). Conversely, below 120 au, the disk is more luminous on its east side than on its west side. This reversed asymmetry might suggest the presence of a spiral density wave extending over the whole disk, and viewed edge-on from the Earth. Finally, a large cavity splits the disk into two debris belts. \cite{Cavite} modeled the stars' excess emission and suggested 13.7 and 46 au for the radii of the belts. The latter likely corresponds to the one imaged by \cite{DisqueSphere} and \cite{DisqueGPI} at $\sim$ 50 au.

Despite the richness of the observations, the geometry and kinematics of the whole system are strongly underconstrained. If the actual planet-binary distance is less than 1,000 au, then the orbit inclination with respect to the plane of the disk must be significant (20 degrees). However, a coplanar configuration cannot be excluded, but the separation should then be around 3,000 au. In any case, the large separation between the planet and the central binary, as well as the possible misalignment between the planet orbit and the debris disk, challenges classical mechanisms of planet formation.

According to current theories, planet formation takes place in the primordial gaseous disk. However, as we mentioned above, forming a giant planet via core accretion or gravitational instability at 700 au or more from any central star appears very unlikely, first due to the lack of circumstellar gas at that distance, and second because the corresponding formation timescale would exceed the lifetime of the gaseous disk. The disk asymmetries (in particular the suspected spiral structure) indicate strong ongoing dynamical interaction with the dust. This may suggest that the planet did not form where it resides today, but may have formed inside and be scattered afterwards. The recently discovered binary nature of HD 106906AB is indeed a source of potentially strong dynamical perturbations that could trigger planet scattering. 

The purpose of this paper is to investigate both analytically and numerically the scenario that could have lead to the present-day characteristics of the HD 106906 system starting from a planet formation within the circumbinary disk. As viscosity-induced migration tends to make the planet move inwards, in Section 2 we will study the likelihood of an ejection via interactions with the binary, and we will then discuss in Section \ref{Stabilization} how the planet could have stabilized on such a wide orbit. Finally, in section \ref{Disk} we briefly
analyze the effect of this scattering scenario on the disk and the processes that could have shaped it as it currently appears.
Numerical simulations in our analysis have been performed using the Swift\_HJS symplectic integration package (Beust \citeyear{Symplectic}), a variant of the Swift package developed by Levison \& Duncan (\citeyear{Swift}), but dedicated to multiple stellar systems.

\section{Ejection}
\label{Ejection}

\subsection{Basic scenario}

We investigated how HD 106906 b, supposed initially orbiting the binary on a nearly coplanar orbit, could have been ejected from the disk via dynamical interactions. When it is located far away enough from its host stars, a circumbinary planet may have a very stable orbit. On the other hand, if it migrates too close to the binary, it undergoes a close encounter with the stars and can be ejected. 

\begin{figure}[h]
        \centering
        \includegraphics[width=\linewidth]{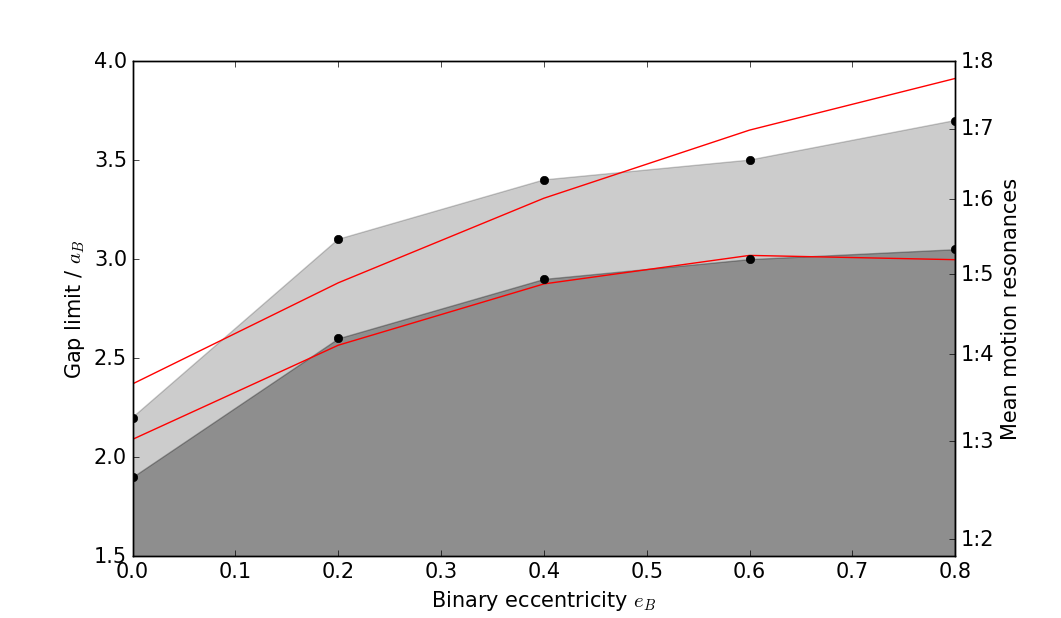}
    \caption{Chaotic zone (in dark gray) as a function of the binary eccentricity, for binary components of same masses. The lighter part designates a critical zone, where some test particles can survive. The red lines represent the lower and upper critical orbit parabolic fits found by \cite{3Corps} in its study of circumbinary planet stability. The 1:6 commensurability is the strongest outside the chaotic zone (see section \ref{SectionMMR}) for $e_B \geq 0.4$.}
    \label{fig:ZC}
\end{figure}

The binary is surrounded by a chaotic zone where no stable circumbinary orbit is possible. \cite{3Corps} uses a semi-analytical approach to compute the upper critical orbit (lower radius of the stable zone) and lower critical orbit (upper radius of the chaotic zone) around two stars of same masses for different binary eccentricities, and found that this gap size typically ranges between two and three times the semi-major axis of the binary orbit. Numerical results for this mass ratio are missing, so that we computed the limits of the gap with our Swift\_ HJS package and compared them to the semi-analytical approach in Fig \ref{fig:ZC}. In each simulation, the evolutions of 10,000 test particles have been studied during $10^5$ orbital periods of the binary. The particles have been randomly chosen with semi-major axes between 1.5 and four times the binary semi-major axis $a_B$, eccentricities between 0 and 0.1, and inclinations with respect to the binary orbital plan between 0 and 3 \degree. The time step has been chosen to be 1/20 of the binary orbital period.

\cite{ZCDisque} showed that this chaotic zone also affects the gas of the disk, with gap sizes similar to the values given by our algorithm \citep{Symplectic}. Consequently, as the migration necessarily stops at the inner edge of the disk, the planet should never reach the chaotic zone this way. It will remain confined close to the lower critical orbit, where it may never be ejected. Mean-motion resonances (hereafter MMR) may help overcoming this difficulty. During its inward migration, the planet is likely to cross MMRs with the binary. It may then be captured by the resonance and furthermore undergo an eccentricity increase that could drive its periastron well inside the chaotic zone. 

\subsection{Mean-motion resonances}
\label{SectionMMR}

Nested orbits are in a configuration of MMR when their orbital periods are commensurable. For fixed masses and neglecting the precession, this is fully controlled by the semi-major axis ratio ${a_B}/{a}$ (subscript B refers to the binary): the orbits are said to be in a $p+q:p$ resonance when 

\begin{equation}
\frac{T_B}{T} = \left(\frac{a_B}{a}\right)^{3/2} \sqrt{\frac{m_B}{m_B+m}} = \frac{p+q}{p}\quad, \label{MMR}
\end{equation}

\noindent where $p$, $q$ are integers, and $T$ and $m$ designate respectively Keplerian periods and masses. Resonances are described using the characteristic angle

\begin{equation}
\sigma = \frac{p+q}{q} \lambda_B - \frac{p}{q}\lambda - \omega\quad,
\end{equation}

\noindent where $\lambda$ designates the mean longitude and $\omega$ the periastron longitude. $\sigma$ represents the longitude of a conjunction between the binary and the planet, where all three bodies are aligned, measured from the line of apsides of the planet. If $\sigma$ stops circulating and begins to oscillate around an equilibrium position (libration), it means that the conjunctions repeatedly occur roughly at the same places on the planet orbit: the system is locked in the resonant configuration.  If the resonant conjunction occurs in the location where the interacting bodies are sufficiently far away from each other (like in the Neptune-Pluto case), then the resonance acts as a stabilizing mechanism that prevents close encounters. MMRs are nevertheless known to enhance eccentricities. If the eccentricities are too highly excited, then the conjunctions may no longer occur at safe locations, often causing instability. For a review on MMRs, see \cite{Celeste}.

\begin{figure}[h]
        \centering
        \includegraphics[width=\linewidth]{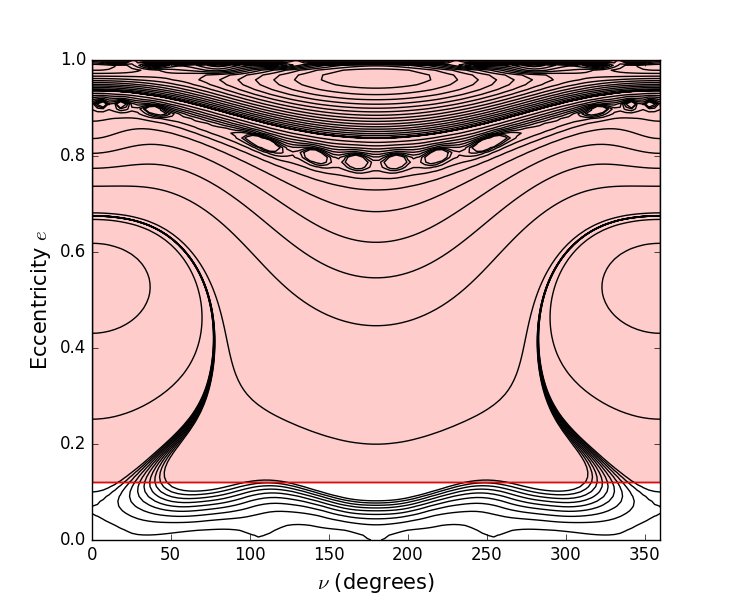}
    \caption{Isocontour in the ($\nu=\omega-\omega_B$,$e$) phase space of the average interaction Hamiltonian of a test particle trapped in 1:6 mean-motion resonance with a binary eccentricity of $e_B=0.4$, assuming a binary mass parameter of $\mu = 1/2$. Each curve represents a trajectory in the ($\nu$,$e$) space. Above the red line, the planet has part of its orbit in the chaotic zone.}
    \label{fig:resonance}
\end{figure}

The way a MMR can enhance the eccentricity of the planet can be studied in a semi-analytical way. Details about this procedure are given in \cite{Planet9}, \cite{BeustMorbidelli} and \cite{Yoshikawa}. Basically, if we restrict the study to orbits with negligible $\sigma$-libration, the interaction Hamiltonian can be averaged over the motion of the binary for constant $\sigma$. This gives a one degree of freedom autonomous Hamiltonian. Phase-space diagrams with level curves of this Hamiltonian can then be drawn in  ($\nu \equiv \omega-\omega_B, e$) space to explore the overall dynamics. To adapt the method to this unusual case where the inner bodies have similar masses, we calculated the resonant Hamiltonian of a planet orbiting the center of mass of a binary with binary mass parameter $\mu \equiv m_2/m_B$ (where $m_2$ is the mass of the second star):

\begin{align}
H = -\frac{Gm_B}{2a} &- Gm_B \left( \frac{1-\mu}{\displaystyle|\vec{r}+\mu\vec{r_B}|} + \frac{\mu}{\displaystyle|\vec{r}-(1-\mu)\vec{r_B}|} - \frac{1}{|\vec{r}|} \right) \nonumber\\
 &- \frac{p+q}{p}\frac{2\pi}{T_B}\sqrt{Gm_Ba}\quad,\label{H}
\end{align}

\noindent where $G$ is the gravitational constant, $\vec{r_B} \equiv \vec{R_2}-\vec{R_1}$ and $\vec{r} \equiv \vec{R} - (\mu\vec{R_1}+(1-\mu)\vec{R_2})$ if $\vec{R}$, $\vec{R_1}$ and $\vec{R_2}$ are respectively the position vectors of the planet, the first and the second component of the binary. We could then perform the integration over the orbital motions and derive the phase space diagram for the interesting commensurabilities. The result is displayed in Fig. \ref{fig:resonance} in the 1:6 MMR case, for a binary eccentricity of $e_B=0.4$. Most of the level curves of the Hamiltonian exhibit important change in the planet eccentricity; therefore, starting at low eccentricity, the resonant interaction can drive the planet to higher eccentricity regime and cause it to cross the chaotic zone (indicated in red on the figure) at periastron, leading to ejection.

Our choice of focusing on the 1:6 mean-motion resonance should not be surprising. Indeed, according to Fig. \ref{fig:ZC}, it is the lowest order resonance that lies outside the chaotic zone for $e_B\geq 0.4$: it occurs at $a/a_B \simeq 3.3$. Any lower order (thus potentially stronger) resonance such as 1:2, 1:3, etc. falls inside the chaotic zone, and could not be reached by the planet according to our scenario. Moreover, the topology of the diagram depends on the binary eccentricity: the higher it is, the higher is the change of eccentricities depicted by the level curves. And those curves are flat for a circular binary orbit.

However, the semi-analytical study is not sufficient here to study the dynamical route that leads to ejection. Indeed, libration of the resonant angle $\sigma$ and chaos on short timescale, not taken into account in the computation of the phase-space diagram, are not negligible for a binary with mass parameter close to 1/2. We thus performed numerical simulations of $10^5$ binary orbital periods of dynamical evolution for different binary eccentricities and different initial angular conditions, to study the stability of different ratio of MMR. All runs were performed starting with a semi-major axis close to the resonant value, with a time step set to 1/20 of the binary orbital period.

Only a few resonances located outside the chaotic zone are finally able to trigger ejection: the 1:6 and the 1:7 one. The simulations allowed to check not only the ability of the resonances to generate instability, but also the time needed to eject the planet, as well as the typical width of the starting resonant zone that leads to ejection, which is typically 0.01 au. Table \ref{table:withoutMigration} summarizes the results obtained with various $e_B$ values and $\mu=1/2$ with the 1:6 resonance. 

\begin{table}
        {\caption{Effect of the 1:6 mean-motion resonance and ejection duration for different eccentricities of the binary, starting with a planet eccentricity of $e=0.05$. For $a_B = 0.4$ au, the 1:6 resonance corresponds to a planet semi-major axis around $a=1.3$ au.}
        \label{table:withoutMigration}                        
        \begin{center}
        \begin{tabular}{c c}        
        \hline\hline\\[-0.8em]                 
        Binary eccentricity & Effect of the 1:6 MMR \\     
        \hline\\ [-0.8em]                          
        $e_B=0.0$ & no ejection \\
        $e_B=0.2$ & no ejection \\
        $e_B=0.4$ & ejection in 100-1000 years  \\
        $e_B=0.6$ & ejection in 100-1000 years  \\
        $e_B=0.8$ & ejection in 100 years  \\ 
        \hline                                   
        \end{tabular}   \end{center}}
\end{table}

The simulations confirm that resonance stability depends on binary eccentricity $e_B$, and that the resonance gets weaker when the order of the resonance $|q|$ increases. An important result is that whenever ejection occurs, it happens within a very short timescale, always much shorter than the typical time needed ($\la 1\,$Myr) to form the planet from the gaseous disk. Our first conclusion is thus that the planet cannot have formed within the resonance. This validates the idea outlined above that it first formed at larger distances in a more stable position, and furthermore migrated inwards and was possibly trapped in a mean-motion resonance before being ejected. In the following, we investigate this scenario.

\subsection{Migration}

In recent decades, planet migration has become an unavoidable ingredient to explain the configuration of some planetary systems. Due to tidal interactions with the primordial gas disk, giant planets (mass > 10 $M_\oplus$) undergo first a type I, and furthermore a type II migration once they have created a gap in the disk \citep{Migration}. It consists of a drift that can be directed toward the star, whose characteristic timescale depends on the position, and characteristics of the planet and on the viscous properties of the disk.

We have assumed that the planet has approximately reached its final mass when it arrives at the location of unstable MMRs, that is between 1 and 2 au from the stars. The characteristic time of migration varies in inverse proportion to the quantity $\alpha_\nu h^2 \Sigma$, where $\alpha_\nu$ is the viscosity parameter, $h$ the aspect ratio and $\Sigma$ the surface density \citep{TypeII}. However, not only the values of those quantities are unknown in HD 106906 primordial disk, but also this simple dependency does not seem to match nor the known planetary population \citep{MigrationCorrection} neither the results inferred by hydrodynamical simulations \citep{TypeIIcorrection}. Taking this fact into account, estimating the mass of the primordial disk to be around 0.6\% of the stellar mass \citep{MasseDisque} and varying the viscosity parameter and the aspect ratio around the observed values (e.g., Pinte et al \citeyear{HLTau}), we obtained a large range of migration timescales. To obtain the largest overview without trying every single velocity, we choose to run our tests with four different migration velocities at 2 au: $10^{-3}$, $10^{-4}$, $10^{-5}$ and $10^{-6}$ au/year.

Simulating the whole process of disk-induced migration in the circumbinary environment is beyond the scope of the present paper. Using a hydrodynamic code, \cite{MigrationBinaire} computed the migration of a planet in a circumbinary disk and show that it was likely to get locked into a mean-motion resonance. As their stars had very different masses, their results can not be applied here, so we choose to add to the SWIFT\_HJS code an additional extra-force that mimics the migration mechanisms they observed. This force is designed in such a way that its secular effect averaged over the orbital motion of the planet just induces the desired steady-state semi-major axis drift $\text{d}a/\text{d}t = v_\text{mig}$, $v_\text{mig}$ being a fixed arbitrary migration velocity, and has no effect on the eccentricity nor on the longitude of periastron. Further details about the choice of the force are provided in Appendix \ref{AppendixMig}. We derive:

\begin{equation}
\vec{F}_\text{mig} = \frac{v_\text{mig} n}{2\sqrt{1-e^2}} \left(1+\frac{1}{2}(1-\frac{r}{a})\right) \vec{e_\theta}\quad,\label{Fmig}
\end{equation}

\noindent where $(\vec{e_r},\vec{e_\theta})$ are the 2-D cylindrical radial and orthoradial unit vectors in the local referential frame attached to the planet's motion. Thus, $\vec{F}_\text{mig}$ depends on the planet position via the radius $r$, the vector $\vec{e_\theta}$ and the planet mean angular motion $n = {2\pi}/{T}$. The parameter $v_\text{mig}$ is set at the beginning of the simulation, according to the timescale we want for the migration. We note that with the above convention, inward migration corresponds to $v_\mathrm{mig}<0$. Of course, the migration is implicitly assumed to hold as long as the planet moves inside the disk.

Whether migration would inhibit or enhance the effects of MMR is not a straightforward issue. Resonance trapping induced by type II migration was found to exist for some commensurabilities between two protoplanets orbiting a star (\citeauthor{MigrationResonance} \citeyear{MigrationResonance}; \citeauthor{MigrationResonance2} \citeyear{MigrationResonance2}). But MMRs with a binary are more difficult to predict, especially those located near the chaotic zone, like those we are focusing on here. Moreover, the expected lifetime of the gas disk is roughly three million years around massive stars (\citeauthor{DiskLifeTime} \citeyear{DiskLifeTime}; \citeauthor{DiskLifeTime2} \citeyear{DiskLifeTime2}), so that the formation, migration and hypothetical ejection must all occur by this time.

\begin{table*}
        {\tiny\begin{center}
        \caption{First unstable resonance and corresponding ejection time for different eccentricities of the binary and different migration velocities, starting with $e=0.05$ and $a=2$ au. The ejection time corresponds to the time needed to eject the planet starting from the beginning of the MMR trapping.} 
        \label{table:withMigration}      
        \centering                         
        \begin{tabular}{l l l l l}        
\hline\hline\\[-0.8em]                 
migration & $10^{-3}$ au/year & $10^{-4}$ au/year & $10^{-5}$ au/year & $10^{-6}$ au/year  \\     
\hline\\[-0.8em]                     
   $e_B=0$ & $1:4$, 50 yr & $1:4$ , $10^3$ yr & $1:5$ , $10^4$ yr & $1:5$ , $10^5$ yr \\
   $e_B=0.2$ & $1:5$, 50 yr & $1:6$ , $10^3$ yr & $1:6$ , $10^4$ yr & $1:7$ , $5.10^4$ yr \\
   $e_B=0.4$ & $1:6$, 100 yr & $1:6$ , 500 yr & $1:6$ , $2.10^3$ yr & $1:7$ , $2.10^4$ yr \\
   $e_B=0.6$ & $1:6$, 100 yr & $1:6$ , 100 yr & $1:6$ , $2.10^3$ yr & $1:6$ , $10^4$ yr \\ 
   $e_B=0.8$ & $1:6$, 100 yr & $1:7$ , 500 yr & $1:8$ , $10^4$ yr & $1:7$ , $2.10^4$ yr \\ 
\hline                                   
\end{tabular}\\
\end{center}}
\end{table*}

We thus performed numerical simulations, where the planet was initially put outside ($\sim$ 2 au) the zone of interest. Whether the planet formed just outside the critical zone or whether it migrated toward there is irrelevant, only the values of the orbital elements and migration velocity at the entrance of the zone of interest matter to conclude on the possibility of ejection. Migration was added using the additional force depicted in Eq. \ref{Fmig}, with the diverse migration velocity prescriptions described above. The simulations were pursued until the planet gets captured in a mean-motion resonance and furthermore ejected, or until it reaches the inner edge of the disk, that is, the chaotic zone, in the case of no resonant capture. Again, the time step has been taken to be $\lesssim T_B/20$. The main result is that migration, regardless of its velocity or of the binary eccentricity, always leads to a resonant trapping followed by an ejection after a reasonable amount of time spent in the resonance. 

\begin{figure}[h]
        \centering
        \includegraphics[width=\linewidth]{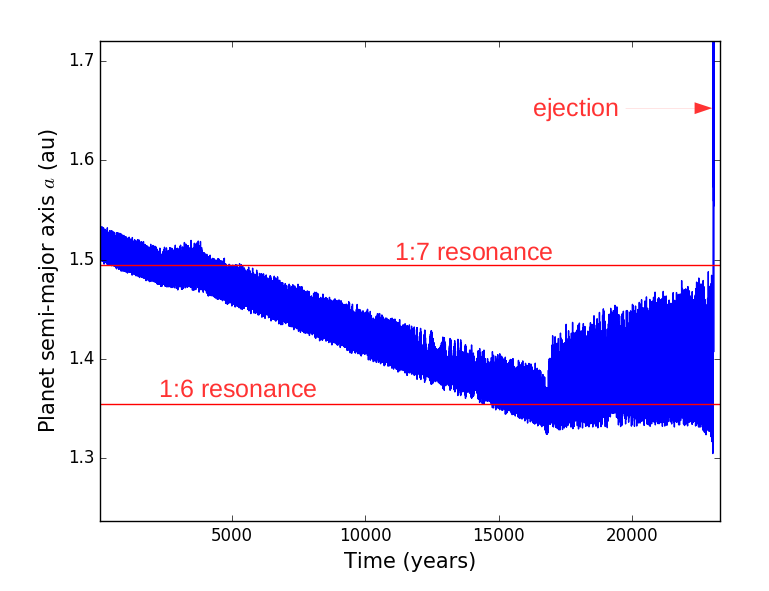}
    \caption{Evolution of the planet semi-major axis with respect to time for a binary eccentricity of $e_B = 0.2$ and a $10^{-5}$ au/year migration velocity. The semi-major axis of the binary has been set to $a_B = 0.4$ au. The plot illustrates the migration, then ejection, of the planet after it has been trapped into a 1:6 resonance. The effect of the 1:7 resonance, weaker, is also visible on the plot. As the planet has a perturbed Keplerian motion around the binary, the exact locations of MMRs are not straightforward to derive (see Appendix \ref{AppendixPrecession}).}
    \label{fig:ejectionMigration}
\end{figure}

In Fig. \ref{fig:ejectionMigration}, an example of the effect of both migration and resonances is visible via the evolution of the semi-major axis of the planet. The figure illustrates the full dynamical evolution corresponding to $v_\text{mig}=10^{-5}$ au/year and $e_B = 0.2$. In addition to high frequency oscillations that illustrate the chaotic nature of the dynamics, we see a gradual semi-major decrease at a speed corresponding to the initial prescription, followed by a capture in the 1:6 MMR resonance that finally leads to ejection. Interestingly, we note a temporary trapping in the 1:7 resonance than occurs before the final capture in the 1:6. The 1:7 resonance appears not to be strong enough to fully inhibit the migration, while the 1:6 does. 

Table \ref{table:withMigration} summarizes the ejection times obtained in the various cases tested. Comparing Tables \ref{table:withMigration} and \ref{table:withoutMigration}, we note that migration, despite causing important small-scale variability of the semi-major axis, enhances resonant instabilities. However, this efficiency is probably overestimated because of the simplicity of our migration model. Deeper analysis of the disk-planet interaction close to the resonance would be needed. Moreover, close to its inner edge, the disk is strongly shaped by the binary and some eccentric ring-like features may affect the protoplanet migration \citep{DisqueSelfGravity}.

We may now summarize the analysis that has been conducted in this section by reviewing the time evolution of this tentative ejection process. The formation of a giant planet takes a variable amount of time depending on the process and the location: from several periods if formed via gravitational instabilities to a million periods if formed via core accretion \citep{Formation}. Consequently, in order for HD~106906~b to acquire its mass, it must have formed in a relatively stable location over the timescale involved, at least at a distance of 2 au. However, as giant planets are believed to form beyond the snow line, whose location is estimated to $\sim$ 10 au around $\sim 3 M_\odot$ star \citep{SnowLine}, the stability of the planet formation position is a priori ensured. After a substantial growth of the planet, migration occurs, whose strength depends on the primordial disk characteristics, and pushes the planet into a less stable zone. For the planet to be ejected, it has to enter the zone of destabilizing resonances (1:6, 1:7), which lies around 1.5 au (Fig. \ref{fig:ejectionMigration}). All in all, if $a_\text{formation}/v_\text{mig}$ is inferior to the disk life-expectancy, the scattering of the planet is a natural outcome in a system with binary mass ratio close to unity.

\section{Stabilization}
\label{Stabilization}

\subsection{The idea of a close encounter}

In the previous subsection, we demonstrate that a giant planet which formed reasonably close to the binary is likely to undergo an ejection. However, ejection does not imply stabilization on a distant orbit around the binary, as it is most likely the case for HD~106906~b. Eventually, the planet follows an hyperbolic trajectory and does not need more than 10,000~years to completely fly away from its host star. Indeed, suppose that the planet gets ejected on a still-bound orbit via a close encounter with the binary: the orbit may have a very distant apoastron, but its periastron will necessarily lies in the region where it originates, that is the immediate vicinity of the binary. Therefore, after one orbital period, the planet is back at periastron and undergoes a new violent encounter with the binary that is likely to cause ejection. Such episodes have been actually recorded in our simulations.

Thus, in order to stop the ejection process and stabilize the planet orbit, an additional dynamical process is needed to lower its eccentricity and increase its periastron. In the absence of other wide companion of similar mass orbiting the binary \citep{DisqueSphere}, a close encounter with a passing star is a natural candidate. Recalling that the Sco-Cen association must have had a more important stellar density several million years ago, this event might have occurred with non negligible probability.

The impact of dynamical interactions on planetary systems in open clusters has been studied intensively since the discovery of the first exoplanets. An effective cross section has been computed by \cite{CrossSection}, that characterizes the minimal encounter distance needed to raise the eccentricity of a Jovian planet at 5 au from $0$ to over $0.5$. They found $\langle\sigma\rangle = (230 \text{ au})^2$, which gives a stellar encounter rate of about 0.01 disruptive encounter in our system lifetime. More precisely, \cite{Parker} conducted N-body simulation to observe the planet orbital elements after a fly-by, and found a probability between 20 and 25\% that a 30 au planet undergoes at least a 10\% eccentricity change in a ten million year period. In our case, the situation seems even easier, because we want to modify the orbit of an unstable planet already far from its star, thus with a trajectory that can easily be swayed. However, the encounter needs to happen at the right time of the planet life, during the $\sim$ 1000 years that would last the ejection. Moreover, the encounter should be weak enough not to definitely eject the planet, but strong enough to circularize the orbit to a reasonable eccentricity. We note that weak encounters are more likely to occur than strong ones.

\subsection{Probability of a stabilizing fly-by}

\begin{figure}[h]
        \centering
        \includegraphics[width=\linewidth]{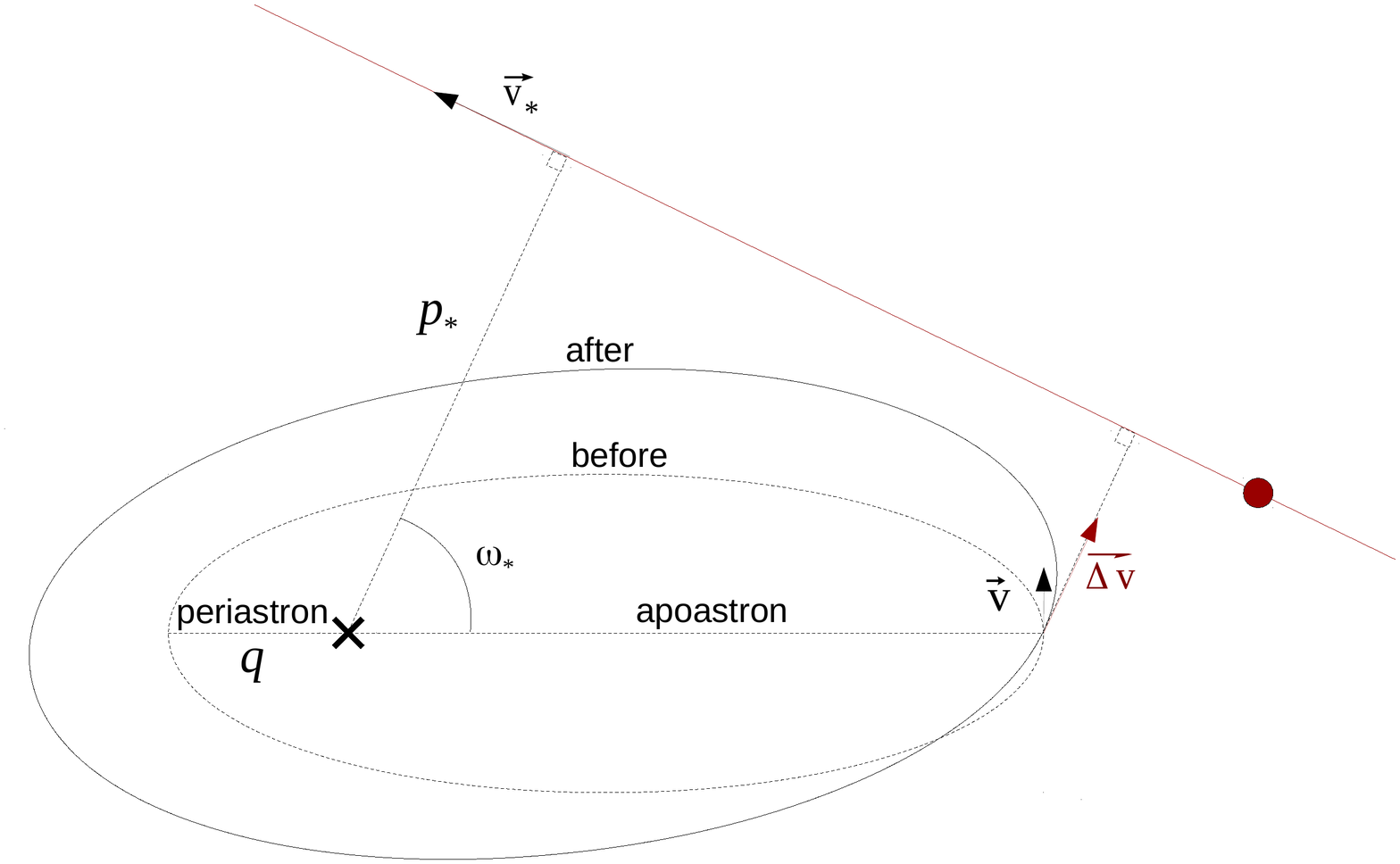}
    \caption{Example of a coplanar configuration where a passing star (in red) stabilizes a wide unstable planet orbit. Before the fly-by, the planet orbit still has a very low periastron, and after it gets much wider, thanks to the interaction with the passing star. We recall that according to Kepler's laws, the planet spends most of its time near apoastron, so that any fly-by is likely to occur when the planet is at or near this point.}
    \label{fig:schema}
\end{figure}

Of course, not all fly-by geometries will generate the desired effect. The fly-by is entirely defined by the mass of the passing star $M_*$, the closest approach (or periastron) to the binary $p_*$, the velocity of the passing star at closest approach $v_*$, the inclination $i_*$ of the passing star orbit with respect to the planet orbit, its longitude of ascending node $\Omega_*$ measured from the line of apsides of the planet orbit, and the argument of periastron $\omega_*$ with respect to the line of nodes. A scheme of the effect of a stellar fly-by is sketched in Fig. \ref{fig:schema} in the coplanar case, in a configuration voluntarily favorable to a restabilization: when the planet is at the apoastron of a wide unstable orbit. In fact, the apoastron is also the most likely position of the planet, as it spend there most of its time.

\begin{figure}[h]
        \centering
        \includegraphics[width=\linewidth]{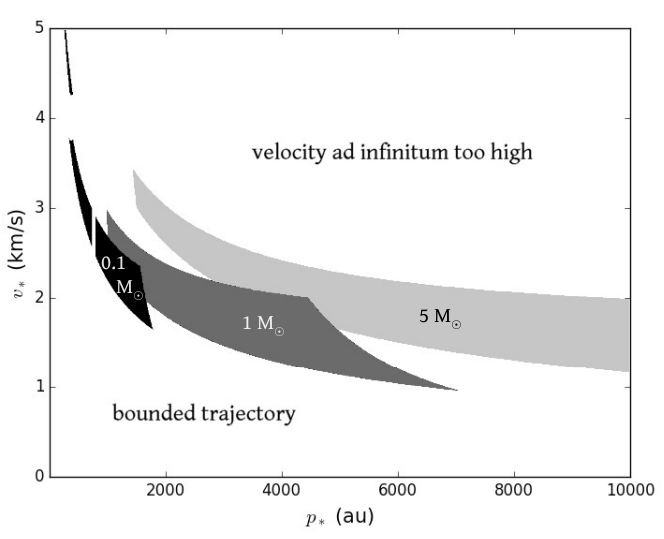}
    \caption{Area in the disruptive star phase space which succeed to raise the periastron from 1 au to over 2 au in the case of a coplanar encounter of periastron argument $\omega_*-\omega = 45$ degrees. $v_*$ designates the maximum relative velocity of the disruptive star, $p_*$ designates its smaller distance from the binary.}
    \label{fig:stabilization}
\end{figure}

Figure~\ref{fig:stabilization} shows the results of a parametric study limited to coplanar fly-bys (we studied the inclined cases as well) for a given angle $\omega_*$ (45 degrees), in $(p_*,v_*)$ 2D parameter space, for three different $M_*$ values (0.1, 1 and $5\,M_\odot$) and assuming the planet was at the apoastron of a very wide unstable orbit before the encounter (such as in Fig. \ref{fig:schema}). The planet is 1,000 au away from the binary when the fly-by occurs, this is why a cut-of can be seen around $p_* = 1000\cos(\frac{\pi}{4})$ au. In each case, the gray area represents the zone in parameter space that is reachable (plausible $v_*$) and actually causes a significant periastron increase of the planet. In this peculiar configuration, taking into account the distribution of $p_*$ and $v_*$ (see below), a stabilization is very likely.

As a more general approach, the probability of a convenient encounter can be estimated by an integration over the relevant fly-by parameters. Taking a homogeneous distribution of stars in the cluster with characteristic distance $d$, and a Gaussian distribution of relative velocities with dispersion $\sigma$, the number of fly-bys that would raise the planet periastron above $q_\text{stable}$ is

\begin{align}
&N_{q_\text{f} > q_\text{stable}} = \frac{\tau_\text{ejection}}{\tau_\text{cluster}} \int_{0}^{+\infty} \frac{dp_*}{4\pi d^2} \int_{0}^{2\pi} p_*d\omega_* \int_{\sqrt{\frac{2GM}{p_*}}}^{+\infty} dv_*~n_v(v_*,p_*) ~ \times\nonumber \\ & \int_{0}^{\pi} di_* \sin i_* \int_{0}^{2\pi} d\Omega_*~ \mathcal{H}e(q_\text{f} - q_\text{stable})\quad,\label{proba}
\end{align} 

\noindent where $q_\text{f}$ is the final periastron reached by the planet after the fly-by perturbation, $\tau_\text{ejection}$ is the characteristic time of ejection, $\tau_\text{cluster} \equiv d/\sigma$ is the characteristic time in the cluster (timescale needed to have a convenient fly-by), $\mathcal{H}e$ is the Heaviside function and $n_v$ the velocity distribution of unbound stars

\begin{equation}
n_v(v_*,p_*) = 4\pi v_* \sqrt{v_*^2 - \frac{2GM}{p_*}} \left(\frac{3}{2\pi\sigma^2}\right)^{3/2} \exp(-\frac{3(v_*^2-\frac{2GM}{p_*})}{2\sigma^2})\quad.
\end{equation} 

This gives the probability of having a stabilizing fly-by, for a given mass $M_*$ of the passing star. Apart from the role of $M_*$ (see Fig. \ref{fig:stabilization}), this probability is strongly though indirectly dependent on the orbital parameters of the planet before the fly-by, that is on the state of advancement of the planet ejection. It is higher when the planet lies initially on a wide, unstable but still bound orbit (as in Fig. \ref{fig:schema}). On the other hand, it is nearly zero as long as the planet is still close to the binary (i.e., before ejection) and if it is already on a hyperbolic trajectory. In order to compute analytically the value of $\mathcal{H}e(q_\text{f} - q_\text{stable})$ for every set of parameters $(p_*,v_*,i_*,\Omega_*,\omega_*)$ given any initial planet position and velocity, we assume a linear trajectory for the perturber. The direction of the velocity change caused by this approximated encounter can then be analytically derived, as well as the new planet orbit . In the computation, we assumed a velocity dispersion of $\sigma = 0.2$ au/year (1 km/s) \citep{Sigma}, and the order of magnitude of the characteristic time of ejection $\tau_\text{ejection}$ has been set to $10^3$ years.

The most critical dependence of our formula (\ref{proba}) is on the local distance between stars $d$. The present and past density of the LCC is not known. Therefore, we attempted to determine it  through a kinematic study in Appendix \ref{AppendixDensity}. From 141 stars for which complete data could be retrieved, we could trace back the density of LCC through time. The results show that the early density was roughly 1.7 times the present density, evaluated around $0.05~\text{star}/\text{pc}^3$ in the close neighborhood of HD~106906. Moreover, the contribution of field stars (not related to LCC) has been estimated to be similar to the contribution of LCC. From this piece of information, we derived that the present local density is lower than $\approx 0.11 ~\text{star}/\text{pc}^3$. This density, consistent with the density of the solar neighborhood \citep{Voisinage}, corresponds to $d\sim 2$ pc. If our scenario happened in such an environment, the probability of a close encounter ($p_*<5000$ au) just following the planet ejection is below $1.10^{-7}$.  

Nevertheless, our estimate of the LCC density is based on a small number of luminous (and mostly early-type) stars for which the kinematics can be inferred. In our case, the fly-by of any object more massive than the planet can stabilize the orbit and impact our probabilities. Therefore, we considered the extreme case where neighboring bodies in the cluster are separated by $d=0.1$ pc, a density similar to the one taken in \cite{CrossSection} and \cite{Parker}. We report the probabilities for that high density and for the case of a 1 $M_{\odot}$ perturber in Table \ref{table:proba}, for different initial conditions. We note that the number of encounters for any $d>5000$ au roughly scale with $d^{-3}$, so that lower-density results can be easily retrieved from the table.

\begin{table}[h]
        \centering  
        {\tiny\caption{Number of close encounters with a 1 $M_\odot$ star raising the planet periastron above a given value (2, 50 or 150 au) depending on the trajectory of the planet before the fly-by. These values have been obtained from Eq. (\ref{proba}).} 
        \label{table:proba}                             
        \begin{tabular}{l l l l}        
\hline\hline                 
Periastron superior to & 2 ua & 50 ua & 150 ua  \\     
\hline\\[-0.8em]                  
   Unstable elliptic trajectory & 8 $10^{-4}$ & 5 $10^{-5}$ & 2 $10^{-5}$ \\
   Slow hyperbolic trajectory & 1 $10^{-4}$ & 2 $10^{-5}$ & 6 $10^{-6}$ \\
   Fast hyperbolic trajectory & < 1 $10^{-6}$ & < 1 $10^{-6}$ & < 1 $10^{-6}$  \\
\hline                                   
\end{tabular}}
\end{table} 

\begin{figure*}
        \centering
        \begin{minipage}[b]{\linewidth}
        \subfloat[]{\includegraphics[width=0.33\linewidth]{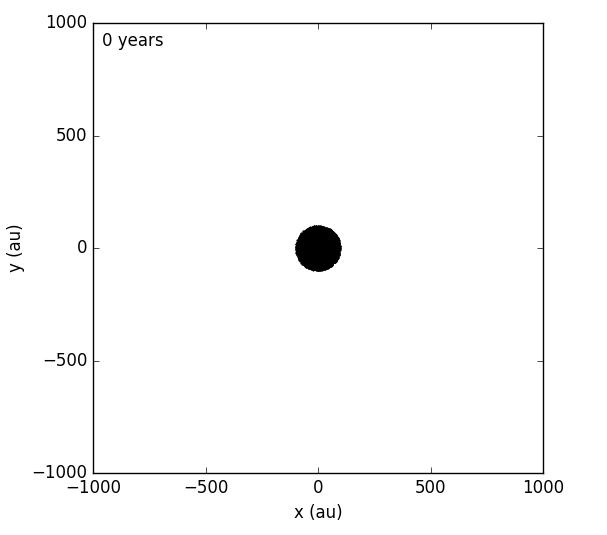}}
        \subfloat[]{\includegraphics[width=0.33\linewidth]{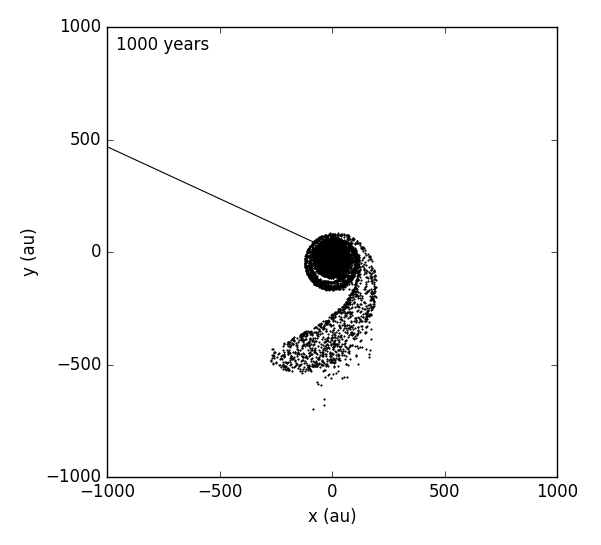}}
        \subfloat[]{\includegraphics[width=0.33\linewidth]{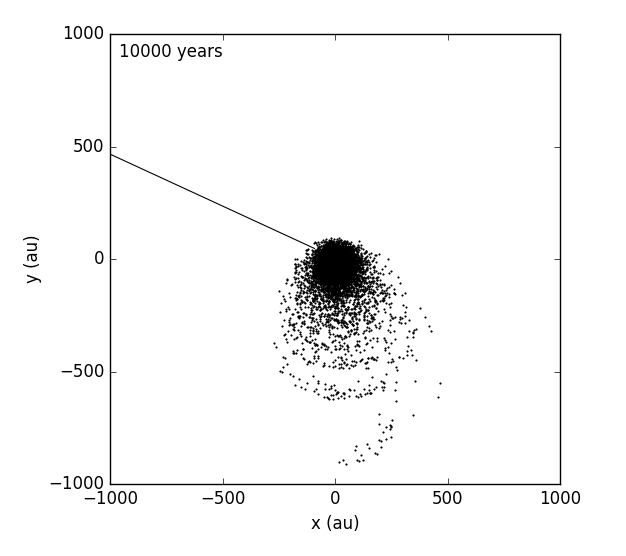}}
\end{minipage}
\begin{minipage}[b]{\linewidth}
        \subfloat[]{\includegraphics[width=0.33\linewidth]{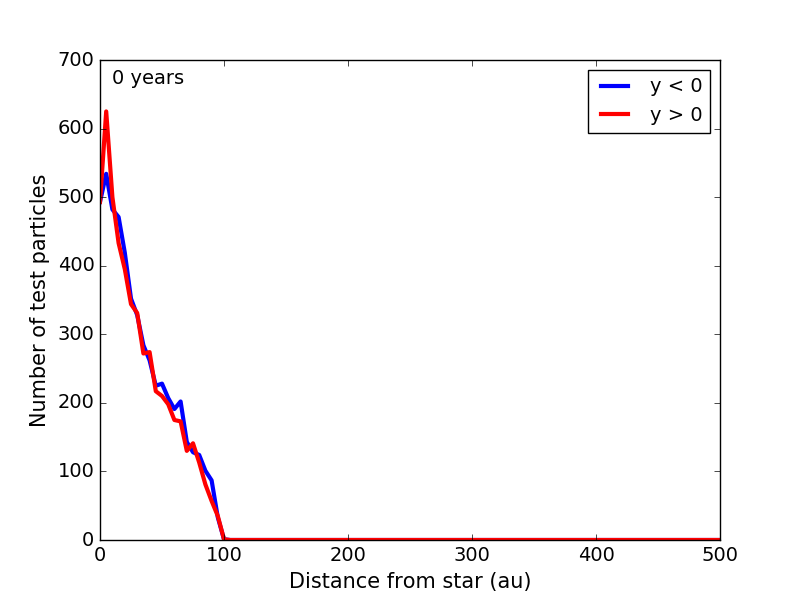}}
        \subfloat[]{\includegraphics[width=0.33\linewidth]{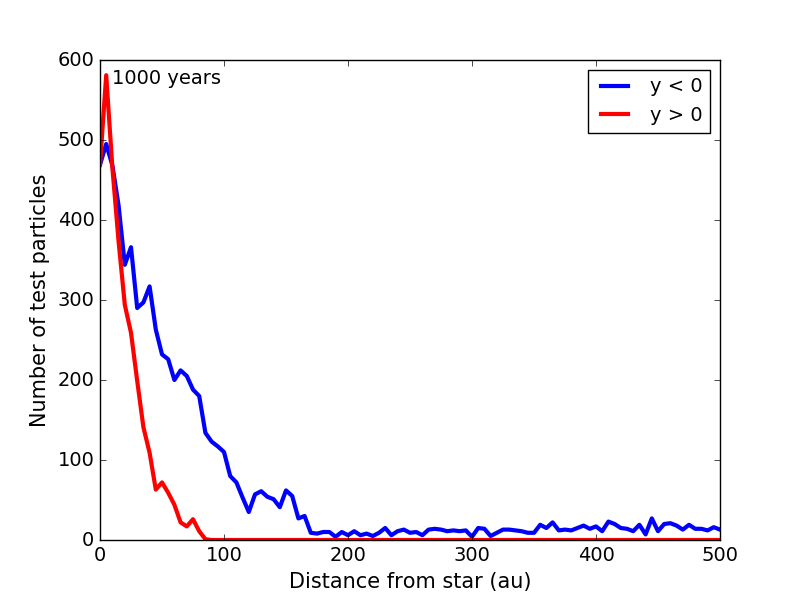}}
        \subfloat[]{\includegraphics[width=0.33\linewidth]{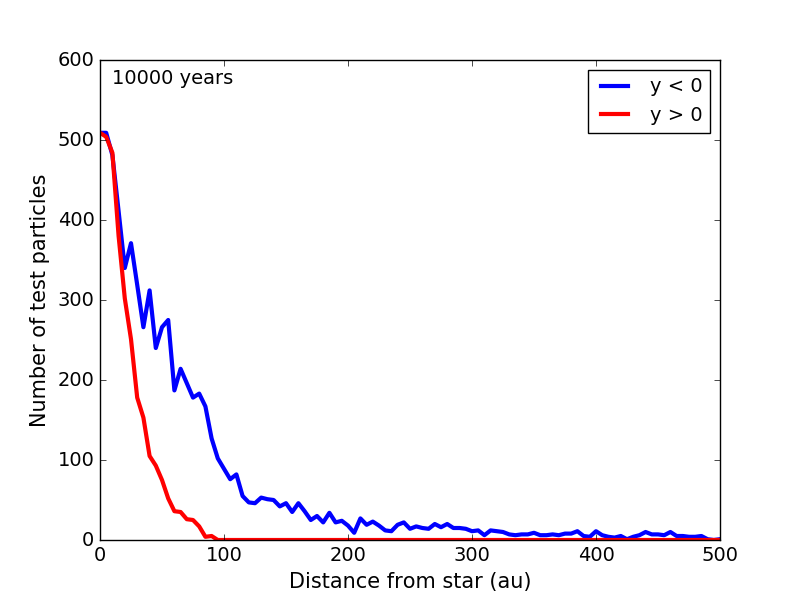}}
\end{minipage}
        \caption{N-body simulation showing the consequence of the ejection of a 11 $M_J$ planet through a disk. From left to right, the snapshots have been taken 0, 1000 and 10,000 years after the ejection. The planet starts on an hyperbolic orbit similar to what we observe in the simulations we performed: $a = 10$ au and $e = 1.1$ (corresponding to periastron $q = 1$ au). Above is a spatial representation of the top view of the disk (the planet trajectory is depicted in black), below is the density along the y axis, integrated over the x and z axis.}\label{fig:ejectionDisk}
\end{figure*}

\subsection{Conclusions}

Table \ref{table:proba} shows that the probability of a stabilizing fly-by remains low. As outlined above, the most favorable case corresponds to initially wide elliptical orbits before encounter. However, most of the time, the planet is directly ejected on an hyperbolic orbit instead of a wide elliptical orbit. And even if this occurs, the subsequent periastron passages in the vicinity of the binary quickly lead to a definitive ejection.

The probabilities have been computed for a 1 $M_\odot$ perturber only, less than half of our system 2.7 $M_\odot$ star. Though the perturber to host star mass ratio do matter to evaluate the fly-by impact \citep[e.g.,][]{Jilkova2016}, the 1 $M_\odot$ results give an upper bound that accounts for the encounter with lighter stars, and a rough estimate for encounters with heavier stars (see Fig. \ref{fig:stabilization}), which are less abundant.

We therefore conclude that while our scenario uses generic ingredients (migration, MMR, fly-by), it is not very likely to happen because of the low probability of a fly-by-assisted stabilization. An indirect proof could be provided if we could see traces of planet ejection on the disk. Moreover, constraints on the present-day orbit of HD 106906 b would certainly help refining this scenario: a very high planet eccentricity could raise its likelihood, but the secular effect of such a planet passage in the disk every thousands of years would have big consequences on the disk morphology.

\section{Debris disk}
\label{Disk}

In this section, we investigate the consequences of our scattering scenario on the disk particles repartition, to check whether it matches the observations (short-distance asymmetry, long-distance asymmetry and extended inner cavity).

\subsection{Ejection through the disk}

An essential part of the scenario we outline in this paper is the violent scattering of the planet by the binary. Most of the time, the planet switches directly from a close orbit around the binary to a fast hyperbolic trajectory toward the edge of the system. As of yet we did not mention the effect of such an ejection on the debris disk surrounding HD~106906AB. The passage of a $\sim 10\,$M$_\mathrm{Jup}$ planet across the disk should presumably induce drastic perturbations on it. In order to investigate this issue, we ran a N-body simulation with 10,000 test particles, neglecting the interactions between them to access the first order of perturbation. The particles have been randomly chosen with semi-major axes between 5 and 100 au, eccentricities between 0 and 0.05, and inclinations with respect to the binary orbital plan between 0 and 2 \degree.  As the main effect of the ejection is due to close encounters between the planet and the disk particles, we use the package Swift\_RMVS \citep{Swift} that is designed to handle such trajectories. However, this package is not devised to work in multiple stellar system, so that the binary will be here approximated by a single star. The binary effect on the dust being negligible above 5 au for the duration of the perturbation (approximately ten times the planet ejection time, that is 10,000 years), this approximation has almost no consequences on the final dust distribution. Time steps have been set to at most 1/20 of the particles orbital periods, but Swift\_RMVS automatically adjusts them to manage close encounters. 

The result is displayed in Fig. \ref{fig:ejectionDisk}. After the initial spiral-like propagation of the eccentricity disturbance created by the planet, the disk homogenizes on an oblong asymmetric shape that could possible match the needle we observe up to $\sim 500\,$au. In the case where the planet is first scattered on a wide eccentric orbit before being ejected, the process gives eccentricity to some test particles, but the effect is negligible compared to the effect of the ejection that comes next. However, in any case, the asymmetry might not last forever. Orbital precession induced by the inner binary (not taken into account in our simulation) should finally randomize the longitudes of periastron of the particles on a much longer timescale and restore the initial axisymmetric disk shape. For a particle orbiting the binary at 100~au, the precession period (see Appendix \ref{AppendixSpiral}) due to the binary is $\sim 4\times10^7$ years. Of course it is shorter closer to the star, but this remains comparable or larger than the age of the system except in the innermost parts of the disk. Hence still observing the asymmetry today at 500~au should not be surprising even if was created a long time ago. However, our mechanism cannot explain the reversed asymmetry at shorter distance. This inversion presumably corresponds to a spiral density wave extending across the disk that needs a steady-state perturbation to be sustained over a long enough timescale.

\subsection{The effect of a stellar encounter}

In Section \ref{Stabilization}, we discussed the possibility that a stellar fly-by could have stabilized the planet mid-ejection. The effect of such encounters on a disk has been studied intensively \citep[for example in][]{FlybyDisque,Jilkova2016}. This effect is of course very dependent on the mass ratio of the stars and on the encounter periastron and eccentricity. 

In fact, most encounters that would stabilize an unstable planet are compatible with the current shape of the disk. We can, for example, consider the case of a 1 $M_\odot$ star perturber. The majority of the suitable encounters have periastrons superior to 1000 au (see Fig. \ref{fig:stabilization}). According to the computations of \cite{Jilkova2016}, this and the high mass of our star implies that all the disk particles will remain bound. Indeed, the transfer radius, that is the minimum radius where capture is possible, is well superior to the observed limit of the debris disk. For our disk to be depleted, the transfer radius should be inferior to $\approx$ 100 au, which corresponds to an encounter periastron around 250 au.
Thus, though the problem is strongly underconstrained, our scenario is likely to be compatible with the existence of the disk.

\subsection{Secular carving}

The secular action of the planet orbiting the binary on its present day large stabilized orbit is an obvious long-term source of perturbation on the disk. We note that we make here a clear distinction between the initial, short term perturbation triggered by the planet on the disk during its ejection process, which effect has been described in the previous subsection, and the long-term secular action of the planet as it moves on its distant bound orbit. It is known that eccentric companions (planets or substellar) orbiting at large distance a star surrounded by a disk create spiral density waves within the disk \citep{Spirales}. To a lesser extent, binaries do the same on circumbinary disks \citep{DisqueSelfGravity}. The following study nevertheless shows that the asymmetry currently observed in the HD 106906 disk cannot be due to the sole action of the binary, but rather requires an outer source of perturbation like the planet, that enhances the density waves induced by the binary.

We investigate here the secular action of the planet on the disk, combined with that of the binary, using simulations with our Swift\_HJS package. Of course with only a projected position, our knowledge of the current orbit of the planet is sparse. Some orbital configurations may nevertheless be ruled out as they would lead to a destruction of the disk. Jilkova and Zwart (\citeyear{DisqueSimulation}) studied intensively the impact of each orbital configuration on the disk via the percentage of particles that remain bound $n_\text{bound}$ and the fraction of bound particles that suit the observation constraints $f_\text{d/b}$. Although nor the disk neither the binary was resolved at that time, their conclusion still can be used, at least on a qualitative level. They showed that a planet periastron larger than 50~au or an inclination larger than five to ten~degrees is enough to keep a relatively good agreement with the observations ($f_\text{d/b}>0.5$) without completely depleting the disk. However, to better match the observations ($f_\text{d/b}>2/3$), the periastron must lie outside the outer radius of the disk. The maximal inclination is constrained by the observation, that is about twenty degrees. No further constraints can be provided by the simulation of Jilkova and Zwart to rule out any inclination between zero and 20 degrees if the planet orbit does not go across the disk. They point out that Kozai-like mechanisms can lead to some wobbles in inclination, but small enough for the disk to remain in a nearly coplanar state.

\begin{figure}[h]
        \centering
        \includegraphics[width=\linewidth]{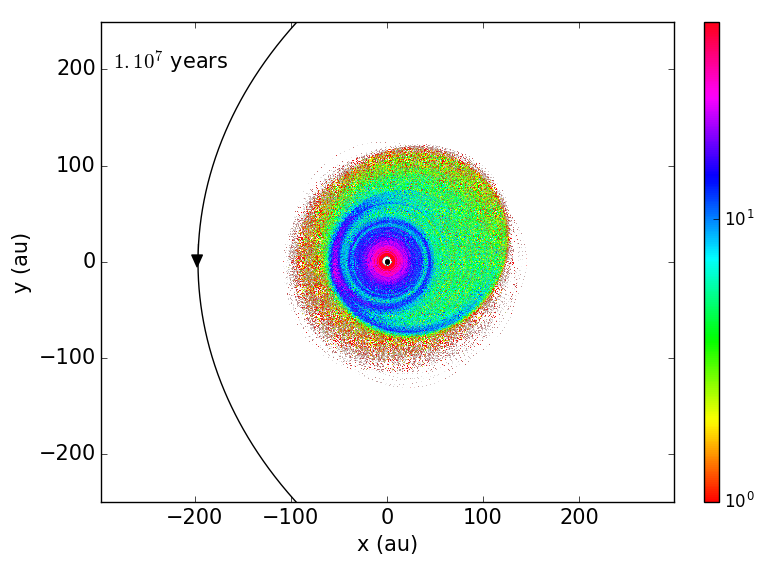}
    \caption{Top view of the evolution of the debris disk after ten million years of perturbation by a planet on a coplanar orbit whose coplanar orbit has a periastron of 200 au and an apoastron of 1000 au. The color scale represents the relative density. Strong asymmetries can be seen.}
    \label{fig:disk}
\end{figure}

\begin{figure}[h]
        \centering
        \includegraphics[scale=0.4]{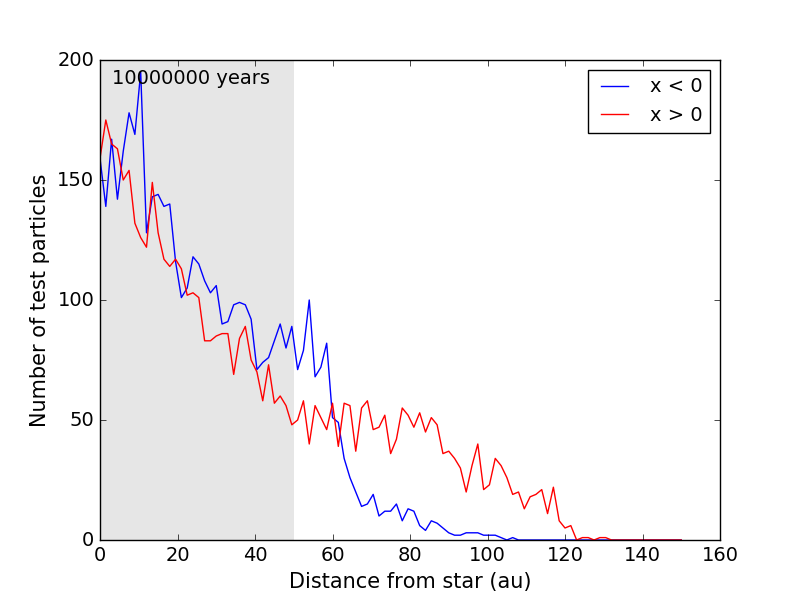}
        \caption{Density along the x axis, integrated over the y and z axis, obtained from Fig. \ref{fig:disk}. The gray zone marks approximately the cavity that we observe today. The asymmetry seems to reverse when we get farther to the stars.}
        \label{fig:Contrast}
\end{figure}

Assuming that the planet fulfills these requirements, we compute the asymmetries induced on the disk and compare the result to the observation. The disk was initially made of 10,000 test particles with same initial conditions than in the previous subsection. The result of a typical run is displayed in Fig. \ref{fig:disk}. Basically, if the periastron of the planet is close enough to the outer edge of the disk, it generates an important asymmetry in the disk within a timescale of between five and ten million years ($\sim 10^8$ binary periods, $\sim 10^3$ planetary orbit). In Fig. \ref{fig:Contrast}, the density profile has been computed along the x axis. The resulting plot displays an asymmetry similar to the observations: the east side (in blue) is brighter than the west side on short scale, but its density drops well above the west side density.  The shape of the perturbation resembles a circular arc, but it actually consists of two overlapping spiral arcs, one driven by the planet and the other one created by the binary.

\begin{figure*}[h]
        \centering
        \includegraphics[width=\linewidth]{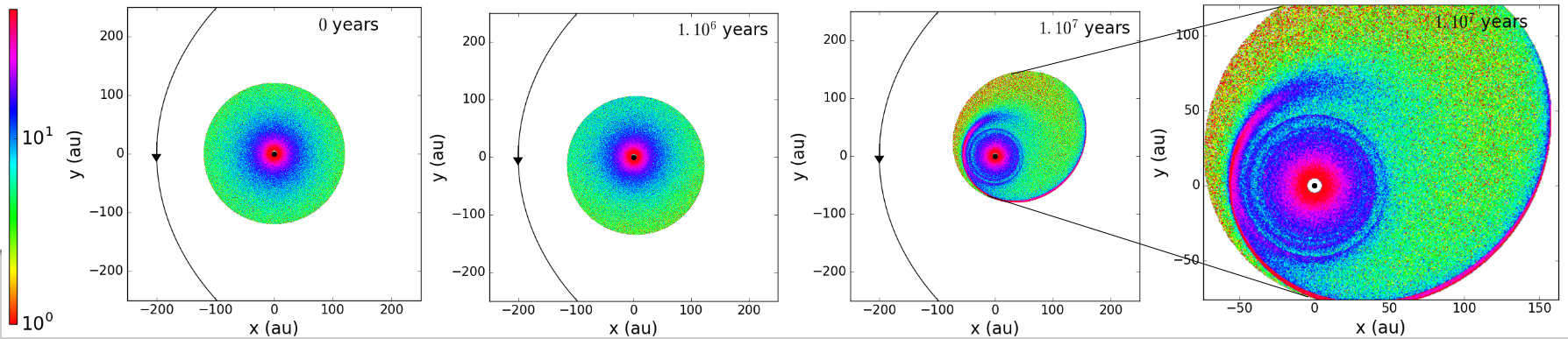}
        \caption{2D representation of a debris disk after 0, $10^6$ and $10^7$ years under the third-order approximation of the influence of the binary and a planet whose periastron is 200 au and apoastron is 1000 au. The color scale represents the local number of particles. At first, only the edge of the disk is affected, but after $10^7$ years, two spirals components appear within the disk.}\label{fig:spirals}
\end{figure*}

This issue can be studied analytically. The approach is analogous to the study without binary, as conducted in \cite{Wyatt}. Consider a test particle orbiting the binary. Suppose that the planet has a Keplerian orbit around the center of mass of the binary, so that the system is hierarchical. The instantaneous Hamiltonian controlling its motion can be written as
\begin{equation}
H=H_\mathrm{Kep}+H_\mathrm{bin}+H_\mathrm{pla}\qquad,
\end{equation}
where $H_\mathrm{Kep}=-Gm_\mathrm{B}/2a$ is the pure Keplerian Hamiltonian, and where the two remaining terms constitute the disturbing function, one part arising from the binary, and the other part from the planet. For a binary of mass parameter $\mu$, these independent perturbations are written

\begin{eqnarray}
H_\mathrm{bin} & = & 
-\frac{Gm_B(1-\mu)}{\left|\vec{r}-\mu\vec{r_B}\right|}
-\frac{Gm_B\mu}{\left|\vec{r}-(1-\mu)\vec{r_B}\right|}-H_\mathrm{Kep}\quad;
\label{hbin}\\
H_\mathrm{pla} & = & -Gm_p\left(\frac{1}{|\vec{r}-\vec{r_p}|}
-\frac{\vec{r}\cdot\vec{r_p}}{r_p^3}\right)\quad,\label{hpla}
\end{eqnarray}
where, in a frame whose origin is at the center of mass of the binary, $\vec{r}$ is the position vector of the particle, $\vec{r_B}$ is the radius vector between the two individual stars, $m_p$ is the mass of the planet, and $\vec{r_p}$ is its position vector. More generally, \emph{B} subscribed quantities will refer to parameters of the binary, \emph{p} subscribed quantities to the planet, while unsubscribed parameters will correspond to the orbiting particle.

Both terms of the disturbing function are then expanded in ascending powers of the semi-major axis ratios $a_B/a$ and $a/a_p$, truncated to some finite order (three here) and averaged independently over all orbital motions, assuming implicitly that the particle is not locked in any mean-motion resonance with the planet or with the binary. Higher orders terms of the disturbing function will be neglected on initial examination, but their influence will be studied in a forthcoming paper. The secular evolution of the particle's orbital elements is then derived via Lagrange equations. Details on this procedure are given in Appendix~\ref{AppendixSpiral}.

Starting from a disk made of particles on circular orbits, we use this theory to compute their instantaneous polar coordinates ($r(t),\theta(t)$) in the disk and compute theoretical synthetic images. The result is shown in Fig. \ref{fig:spirals}, which must be compared with Fig. \ref{fig:disk}. We note the presence of a circular arc very similar to the one obtained in the numerical simulation. This peculiar shape is due to the combination of two spiral waves winded in opposite senses, induced by the planet and the binary via differential precession and eccentricity excitation on the disk particles.  

\begin{figure}[h]
        \centering
        \subfloat{\includegraphics[width=0.9\linewidth]{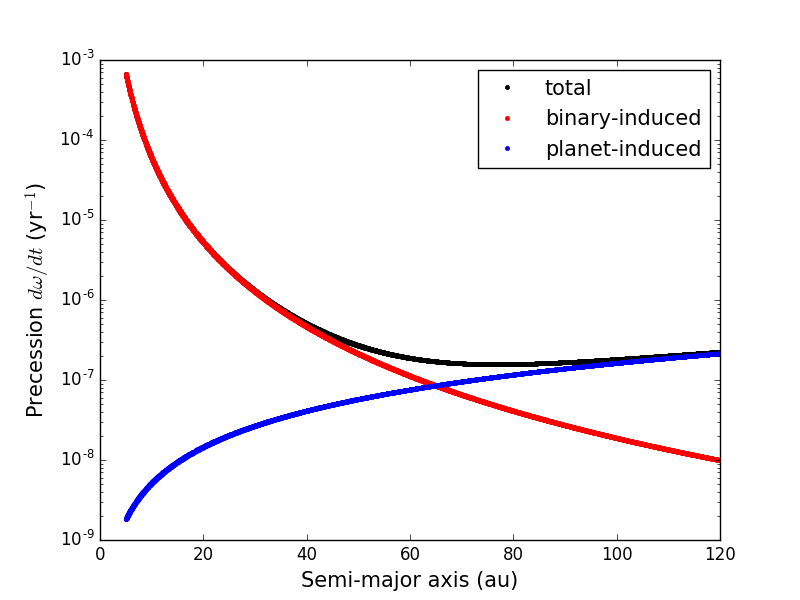}}\\
        \subfloat{\includegraphics[width=0.9\linewidth]{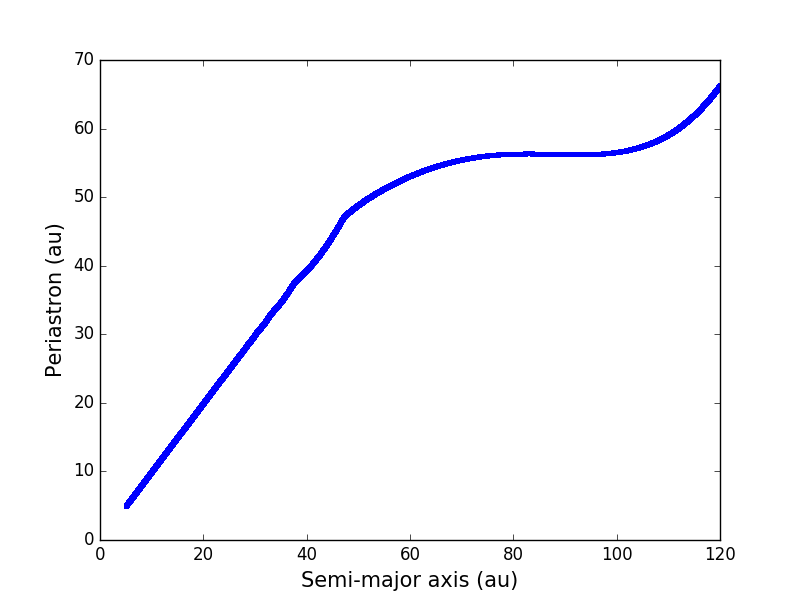}}
\caption{Precession velocities (above) and periastrons (below) of Fig. \ref{fig:spirals} test particles with respect to their semi-major axis (that does not change with time) after ten million years evolution. Above, the red curve displays the precession induced by the binary, the blue curve by the planet and the black curve depicts both contributions.}\label{fig:arc}
\end{figure}

The test particles precession velocities and periastrons are represented in Fig. \ref{fig:arc} as a function of their semi-major axis. In the inner part of the disk, the precession is dominated by the binary, so that the speed of the orbital precession decreases with increasing semi-major axis. The results is a trailing spiral wave that can be seen in Fig. \ref{fig:spirals}. Conversely, in the outer part of the disk, the precession is mostly due to the planet, so that its is now an increasing function of the semi-major axis. This creates a leading spiral density wave. The superposition of both spirals in the intermediate region generates the observed circular arc. The exact location of this arc corresponds to the periastron of the particles whose semi-major axis minimizes the precession velocity, that is, around 55 au. Moreover, we see in Fig. \ref{fig:arc} that all particles from $a \sim 60$ au to $\sim 100$ au have the same periastron. The combination of the two effects enhances the density of the arc, as can be observed in Figs. \ref{fig:disk} and \ref{fig:spirals}.

The two spirals are, however, not fully independent. Orbital precession of the particles actually has no visible effect on their global distribution in the disk as long as their orbits are circular. In Appendix~C, we show that due to the small size of its orbit and its mass ratio close to one, the binary has very little influence on the eccentricity of the particles compared to the planet, even in the inner part of the disk. In fact, while the outer spiral is fully due to the planet, in the inner spiral, the eccentricity oscillations are also driven by the planet, while the precession is controlled by the binary. Moreover, the contrast of the density wave highly depends on the planet orbital shape (the amplitude of the eccentricity oscillations is roughly proportional to $e_p$ within our approximation). For example, if the planet apoastron is 1000 au, its periastron should be less than 500 au ($e > 0.3$) in order to create a significant asymmetry as the one we observe, in a reasonable timescale.

\section{Discussion}

\subsection{Disk cavity}

\begin{figure}[h]
        \centering
        \includegraphics[width=\linewidth]{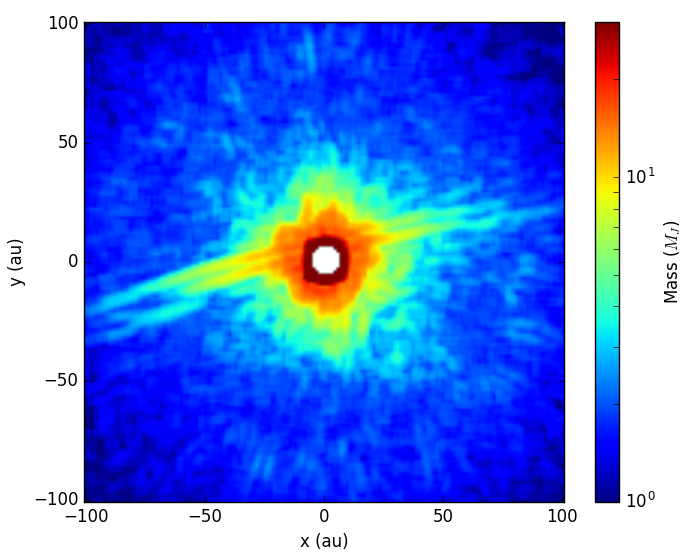}
        \caption{Minimum  mass of planets (in Jupiter masses) that can be detected into the H2 data published in \cite{DisqueSphere} around HD~106906.  The contrasts has been translated into masses using \cite{Modele} model adapted to the SPHERE filters.}
        \label{fig:LimDetection}
\end{figure}

In the previous sections, we did not study the origin of the large cavity observed within the disk by \cite{DisqueSphere} and \cite{DisqueGPI}. It is possible that one or more unseen planets could have carved and sustained this cavity.  In that case,  one of those unseen planet may be responsible for the ejection of the known one, instead of one of the binary star. Those planet(s), if on  eccentric orbits, could also influence the shape of the disk \citep{Morphologies}. Therefore, we ran N-body simulations with the Swift\_RMVS package (the same setting as in the Debris Disk section) to quantify at first order the minimum mass of a single planet, checking if it can carve the observed cavity between the two belts of debris surrounding the pair of stars. This assumes that if one planet alone is responsible for gap, its mass will be higher than in the case where multiple planets are considered. The end result must reproduce the inner edge of the cavity at 13 Myr located between 10 and 15~au \citep[inferred from the IR excess modeling, ][]{Cavite}. We assume that the outer edge of the cavity around corresponds to the separation of the ring (65~au) measured on the SPHERE images \citep{DisqueSphere}. The simulations give a minimum mass of 30~$M_J$ for a single non-eccentric planet located at 30~au, which is well above the detection limits in Fig. 6 of \cite{DisqueSphere}. However, the disk is viewed edge-on, so that the coronagraph used during the observation (radius of 93 mas or 9.5 au) hides part of the orbital plan. In fig. \ref{fig:LimDetection} we computed a 2D detection-limits map from the data published in \cite{DisqueSphere}. The map confirms that a small zone around the coronagraph has detection sensitivity above 30 $M_J$. An additional giant planet on a 30~au circular orbit  will spend  20 \% of its time (ten years) in this blind zone and therefore could have been missed. For the case of an eccentric orbit, the mass of the perturber could be only 1~$M_J$. This is too low compared to the known planet mass to produce an ejection, but high enough to have a noticeable effect on the disk morphology.

\subsection{Alternative scenarios}

Our scenario makes use of standard ingredients (resonances, migration, scattering, fly-by) envisioned or observed in young planetary systems \citep{Migration} and account for all known components of the system. Nevertheless, the low probability of occurrence we estimate in Section \ref{Stabilization} because of the need for a nearby star fly-by at the right time makes the scenario implausible. If we suppose that the planet was in fact on a stable orbit before the fly-by, then this fly-by event could have happened at any time, and not necessarily during the early age of the system. However, the probability for a fly-by to have a significant effect on the planet without ejecting it decreases dramatically when the planet gets closer to its host star. Taking the data from \cite{Parker}, we can expect a probability of around $0.1$ for a disruption superior to 10\% on eccentricity without ejection for a 30 au Jovian planet in the system lifetime. Among the disruptive encounters, it is then hard to tell how many would put the planet on a suitable orbit (apoastron greater than 700 au, that is $e>0.75$). Plus, such a change of orbit would lead to a very small planet periastron, which will strongly deplete the disk (see section \ref{Disk}.2).

Alternatively, the planet could have been stolen from another system. Indeed, captured planets tend to have eccentric orbit \citep{Flyby}. However, for the final orbit to be so wide, the initial orbit must also have been wide \citep{Jilkova2016}. All in all, such a scenario would only turn over the problem, as we would have to account for the wide initial orbit on the first place.

Conversely, the disk could replace the fly-by in our scenario. Indeed, to follow the idea of \cite{Kikuchi}, the planet could have been accelerated by the gas at its apoastron after a first scattering, and its orbit could have been rendered stable this way. It is interesting to note that some of the circumstellar disks of $\sim 2.5 M_{\odot}$ stars recently resolved with ALMA at high angular resolution shows gas extending up beyond the separation of  HD~106906~b (e.g., Walsh et al. \citeyear{GrandDisque}, and ref therein). The total mass of HD~106906 A and B is around 2.7 $M_{\odot}$ and it is therefore possible that the binary bore such an extended primordial disk that would have circularized the orbit of the ejected HD~106906~b. 

\begin{figure}[h]
        \centering
        \includegraphics[width=\linewidth]{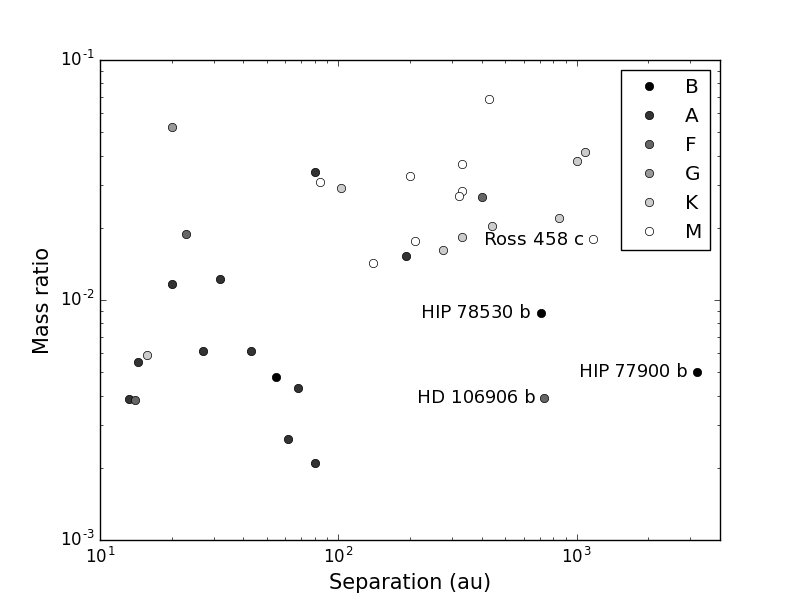}
    \caption{Planet mass ratios with respect to projected separation. Only planets that belong to young systems (< 0.1 Gyr) are displayed, with the exception of the circumbinary planet Ross 458 c. HD 106906 b has the lowest mass ratio beyond 100 au. HIP 78530 b and HIP 77900 b are its two closest neighbors in the diagram. Being identified as brown dwarfs, they may have formed in situ by cloud collapse. Data retrieved from the exoplanets.eu database.}
    \label{fig:Statistics}
\end{figure}

Before the discovery of HD~106906AB binary status that indicates strong gravitational interactions, \cite{Decouverte} suggested that it may have formed in situ. On the one hand, the existence of extended protoplanetary disks implies that our planet may have formed in HD 106906AB primordial disk. On the other hand, HD 106906 b is not the only planetary-mass companion detected at very large projected separation, and such bodies have usually no known scatterers in their environment (see Bryan et al. \citeyear{Scatterers}, even though their study was conducted over a small number of systems less wide than HD 106906 and with lighter stars). Among the systems harboring a planetary-mass companion of similar separation and mass ratio, we can name HIP~77900\footnote{Contrary to HD~106906~b, HIP~77900~b has not been confirmed by the common proper motion test. Nonetheless, \cite{Search} argue that low-gravity features in HIP~77900~b spectrum is compatible with the object being a member of Sco-Cen, and therefore a plausible companion to HIP~77900~A.} \citep{Search}, HIP 78530 \citep{HIP}, or the triple system Ross~458 \citep{Ross}. In Fig. \ref{fig:Statistics}, we represented the wide young planetary-mass companions discovered by direct imaging. We note that HD~106906~b has the lowest planet/star mass ratio above 100 au. The proximity of HIP~78530A~b and HIP~77900A~b  (two brown dwarfs that are also part of Sco-Cen) in that diagram,  could indicate that HD~106906AB~b formed in situ (within the disk, or like a stellar companion). 

\section{Conclusion}

We have shown that HD 106906AB~b could have formed within the primordial disk and be scattered away on a wide orbit during the first ten million years of the system life. This scenario involves the combination of disk-induced migration and mean-motion resonances with the binary. However, if the scattering is likely to occur, the stabilization of the planet on its current wide orbit is delicate, and requires more than gravitational interactions with the binary. A fly-by scenario has been suggested, but the stabilization only occurs for a restricted part of the overall encounters trajectories. The low density (<0.11 stars/pc$^3$) that we estimated for the LCC makes a close encounter even more unlikely.

The disk has multiple features, that each could be explained within the frame of our scenario, but also outside of this frame. Two spiral density waves are created if the planet have for the last ten million years had an eccentric orbit with periastron around 200 au. A needle extending to 500 au could have been created by the ejection of the planet, but a smaller needle could be provoked by an eccentric and inclined outer orbit (see Fig. \ref{fig:disk}) or by an eccentric inner orbit \citep{Morphologies}. \cite{Nesvold} also studied the secular effect of an eccentric, inclined outer orbit for HD 106906~b in a recent paper, and could produce asymmetries whose brightness repartition is consistent with the observations.

The scenario we explored builds on the observed components of the system (disk, binary star) and on the hypothesis that the planet could not have formed via core accretion or gravitational instability at several hundreds of au.  Nevertheless, the low probability of occurrence of our scenario demands that we reconsider those assumptions. Alternative hypothesis like the circularization of the planet orbit via the interaction with the disk gas or in situ formation could explain the present architecture of the system. But this requires that the disk extends up to the separation of the planet and contains enough gas at that separation.  Recent high quality images of circumstellar disks extending beyond 700 au around massive stars and the close properties of other systems in Sco-Cen (HIP~78530A~b and HIP~77900A~b) argue for this alternative formation pathway.

Finally, we note that many of the methods depicted here are easily generalizable to other circumbinary environment. N-body simulations with a simple migration force can be applied on any circumbinary planet to have a quick overlook of the stability of its early trajectory. Fly-by may not be the most efficient process to stabilize a planet, because of the rarity of suitable close encounters. Destabilization by a fly-by is much more probable. Finally, ejection, outer and inner orbits can create huge asymmetries in the disk during the first ten million years of a system. In particular, an inner orbit enhances the dynamical perturbations created by an outer orbit by speeding up the precession, while the outer orbit if eccentric can enhance the perturbations created by the inner orbit by providing eccentricity to the disk. 

\begin{acknowledgements}
We thank the anonymous referee for reviewing our work and for insightful comments which improved the manuscript. The   project   is   supported   by   CNRS,   by   the   Agence Nationale de la Recherche (ANR-14-CE33-0018, GIPSE), the OSUG@2020 labex and the Programme National de  Planétologie  (PNP,  INSU)  and  Programme  National  de  Physique  Stellaire (PNPS,  INSU). Most of the computations presented in this paper were performed using the Froggy platform of the CIMENT infrastructure (https://ciment.ujf-grenoble.fr), which is supported by the Rh\^one-Alpes region (GRANT CPER07\_13 CIRA), the OSUG@2020 labex (reference ANR10 LABX56) and the Equip@Meso project (reference ANR-10-EQPX-29-01) of the programme Investissements d'Avenir, supervised by the Agence Nationale pour la Recherche. This work has made use of data from the European Space Agency (ESA) mission {\it Gaia} (\url{http://www.cosmos.esa.int/gaia}), processed by the {\it Gaia} Data Processing and Analysis Consortium (DPAC,\url{http://www.cosmos.esa.int/web/gaia/dpac/consortium}). Funding for the DPAC has been provided by national institutions, in particular the institutions participating in the {\it Gaia} Multilateral Agreement. This research has made use of the SIMBAD database operated at the CDS (Strasbourg, France). P.A.B.G. acknowledges financial support from the São Paulo State Science Foundation (FAPESP). We thank Cecilia Lazzoni and François Ménard for fruitful discussions. 
\end{acknowledgements}

\bibliographystyle{aa}
\bibliography{Biblio}

\begin{appendix}
\setlength{\parindent}{0pt}

\section{Ad-hoc force to account for type II migration}
\label{AppendixMig}

        We searched for a coplanar migration force $\vec{F}_\text{mig}$ that induces a constant variation of the planet average semi-major axis, but no change in the planet average eccentricity, nor in the periastron longitude. Let $\vec{F}_\text{mig} = F_r \vec{e_r} + F_\theta \vec{e_\theta}$ be the description of the force in the local referential attached to the planet movement. Gauss equations are \citep{MecaCeleste}: 

\begin{align}
        C\frac{da}{dt} &= 2a^2 (F_\theta + e\vec{F}_\text{mig}.\vec{e_y})\quad;\\
        C\frac{de}{dt} &= r (e+\cos\theta)F_{\theta} + a(1-e^2)\vec{F}_\text{mig}.\vec{e_y}\quad;\\
        Ce\frac{d\omega}{dt} &= r \sin\theta F_\theta - a(1-e^2)\vec{F}_\text{mig}.\vec{e_x}\quad,      
\end{align}

where $\vec{e_x}$ and $\vec{e_y}$ are the vectors in the fixed frame and $C = \sqrt{GMa(1-e^2)}$. We want to assume a simple form for $F_r$ and $F_\theta$ that could then be easily averaged over time. The simplest position-dependent force would be $F_r = A (1 + c \cos u)$ et $F_\theta = B (1 + d\cos u)$, where $A$, $B$, $c$ and $d$ are unknown functions of $(a,e)$, constant at first order over a one-period integration. Our conditions are then summarized to:

\begin{align}
        \frac{d\bar{a}}{dt} = v_\text{mig} &\iff \frac{2B\sqrt{1-e^2}}{n} = v_\text{mig}\quad;\\
        \displaystyle \frac{d\bar{e}}{dt} = 0 &\iff B(2e^2d-3e^3+3e-2d) = 0\quad;\\
        \frac{d\bar{\omega}}{dt} = 0 &\iff A\sqrt{1-e^2}(c-2e) = 0\quad,
\end{align}

where $n = \sqrt{GM/a^3}$ is the mean motion. Taking $A = 0$, $B = {nv_\text{mig}}/{(2\sqrt{1-e^2})}$, any $c$ and $d = 3e/2$, we finally obtain Eq. (\ref{Fmig}).

\section{Location of mean-motion resonances}
\label{AppendixPrecession}

Equation (\ref{MMR}) gives the semi-major axis of a resonant circumbinary planet when its orbit is purely Keplerian. When we take into account the perturbation caused by the binary on the planet orbit, the commensurability of periods that characterizes MMRs can not be easily associated with a semi-major axis, mainly due to orbital precession.

The movement of a circumbinary planet (binary of mass parameter $\mu$) is controlled by the Hamiltonian 

\begin{align}
H &= -\frac{Gm_B}{2a} - Gm_B \left( \frac{1-\mu}{\displaystyle|\vec{r}+\mu\vec{r_B}|} + \frac{\mu}{\displaystyle|\vec{r}-(1-\mu)\vec{r_B}|} - \frac{1}{|\vec{r}|} \right)\\
&= H_\text{Kep} + H_\text{bin}\quad,
\end{align}

where $H_\mathrm{Kep}=-Gm_\mathrm{B}/(2a)$ is the Keplerian Hamiltonian, and where $H_\mathrm{bin}$ is given by Eq.~(\ref{hbin}). If the planet orbits at sufficiently large distance from the binary, $H_\mathrm{bin}$ is a perturbative term that triggers orbital evolution of the planet. This can be investigated analytically via a truncated expansion of $H_\mathrm{bin}$ in ascending powers of $a_B/a$, and an averaging over both orbital motions. To lowest order, this yields

\begin{equation}
H_\mathrm{bin}\simeq-\frac{\mu(1-\mu)}{4}\frac{Gm_Ba_B^2}{a^3}
\frac{3e_B^2+2}{(1-e^2)^{3/2}}\quad.
\end{equation}

Strictly speaking, this approximation is not valid at the exact location of MMRs, as the motions of both orbits are not independent anymore, but it gives a good insight of the perturbation of the planet orbit when it is near the MMRs. Moreover, numerical verifications show that this order two approximation is still relevant for $a \geq 3a_B$, and could thus be made to study the 1:6 resonance. Lagrange equations \citep{MecaCeleste} then give

\begin{align}
\frac{d\omega}{dt} &= \frac{3\mu(1-\mu)}{4} n \left(\frac{a_B}{a}\right)^2 \frac{\frac{3}{2}e_B^2 + 1}{(1-e^2)^2}\quad;\\
\frac{d\lambda}{dt} &= n + \frac{3\mu(1-\mu)}{2} n \left(\frac{a_B}{a}\right)^2 \frac{\frac{3}{2}e_B^2 + 1}{(1-e^2)^\frac{3}{2}} +\left(1 - \sqrt{1-e^2}\right)\frac{d\omega}{dt}\quad,
\end{align}

where $n$ is the Keplerian mean-motion. Thus, if $T_0(a)$ is the Keplerian period $2\pi/n$, then the period of the mean longitude $T_\lambda$ is

\begin{equation}
T_\lambda = \frac{T_0(a)}{\displaystyle 1+\frac{3\mu(1-\mu)}{4}\left(\frac{a_B}{a}\right)^2 \frac{\frac{3}{2}e_B^2 + 1}{(1-e^2)^\frac{3}{2}}\left(1+\frac{1}{\sqrt{1-e^2}}\right)}\quad.
\end{equation}

The MMR configuration is characterized by the steadiness of $\sigma = ({p+q})/{q} \lambda_B - {p}/{q}\lambda - \omega$. However, in our study, the planet orbit remains almost circular until ejection, so that the planet line of apsides is not a good reference. Taking the binary line of apsides (constant in time) as the new reference, the resonance characterization writes $T_\lambda = {p}/({p+q}) T_B$. All in all, the resonant location $a_\text{res}$ satisfies

\begin{equation}
T_\lambda(a_\text{res}) = \frac{p}{p+q} T_B\quad.
\end{equation}

\section{Spiral density wave}
\label{AppendixSpiral}

As mentioned in the text, the motion of a particle moving in a circumbinary disk is controlled by the Hamiltonian $H_\mathrm{Kep}+H_\mathrm{bin}+H_\mathrm{pla}$, where $H_\mathrm{bin}$ and $H_\mathrm{pla}$ are perturbative terms given by equations (\ref{hbin}) and (\ref{hpla}). Following the approach of \cite{Wyatt}, these terms are then expanded in ascending powers of $a_B/a$ and $a/a_p$, truncated to some finite order and averaged over the orbital motion of both orbits (see \cite{Laskar}). To second order and third order, the result is

\begin{align}
U_2 = &-\frac{\mu(1-\mu)}{4} \frac{Gm_Ba_B^2}{a^3} \frac{\frac{3}{2} e_B^2 + 1}{(1-e^2)^\frac{3}{2}} -\frac{1}{4} \frac{Gm_pa^2}{a_p^3} \frac{\frac{3}{2} e^2 + 1}{(1-e_p^2)^\frac{3}{2}}\quad;\\
U_3 = &\frac{15\mu(1-\mu)(1-2\mu)}{16} \frac{Gm_Ba_B^3}{a^4} \frac{e\cos(\omega-\omega_B)e_B(\frac{3}{4}e_B^2 + 1)}{(1-e^2)^\frac{5}{2}}\nonumber\\
&+ \frac{15}{16} \frac{Gm_pa^3}{a_p^4} \frac{e_p\cos(\omega-\omega_p)e(\frac{3}{4}e^2 + 1)}{(1-e_p^2)^\frac{5}{2}}\quad.
\end{align}

In HD~106906 configuration, the binary mass parameter is very close to $1/2$, and the semi-major axis of the binary is very small compared to the distance between the binary and the relevant part of the disk, between 50 and 100 au. Thus, the binary part of $U_3$ can be neglected. Using Lagrange equations, we derive the equations of evolution:

\begin{align}
\frac{d\omega}{dt} =& \frac{3}{16} n \left(\frac{a_B}{a}\right)^2 \frac{\frac{3}{2}e_B^2 + 1}{(1-e^2)^2} + \frac{3}{4} \frac{m_p}{m_B} n \left(\frac{a}{a_p}\right)^3 \frac{\sqrt{1-e^2}}{(1-e_p^2)^\frac{3}{2}}\nonumber\\ &- \frac{15}{16} \frac{m_p}{m_B} n e_p \left(\frac{a}{a_p}\right)^4 \frac{\sqrt{1-e^2} \cos(\omega-\omega_p)(1+\frac{9}{4}e^2)}{e (1-e_p^2)^\frac{5}{2}}\quad;\\
\frac{de}{dt} =& - \frac{15}{16} \frac{m_p}{m_B} n e_p \left(\frac{a}{a_p}\right)^4 \frac{\sin(\omega-\omega_p)}{(1-e_p^2)^\frac{5}{2}}\quad;\label{dedt}\\
\frac{dM}{dt} =& \frac{3}{16} n \left(\frac{a_B}{a}\right)^2 \frac{\frac{3}{2}e_B^2 + 1}{(1-e^2)^2} - \frac{7}{4} \frac{m_p}{m_B} n \left(\frac{a}{a_p}\right)^3 \frac{\left(1+\frac{3}{7}e^2\right)}{(1-e_p^2)^\frac{3}{2}}\nonumber\\ &+ \frac{15}{16} \frac{m_p}{m_B} n e_p \left(\frac{a}{a_p}\right)^4 \frac{\cos(\omega-\omega_p)(1+\frac{29}{4}e^2+\frac{9}{4}e^4)}{e(1-e_p^2)^\frac{5}{2}}\quad,\label{dMdt}
\end{align}

where $M$ is the mean anomaly, $n = \sqrt{Gm_B/a^3}$ is the mean motion and $a$ is a constant of motion in the secular regime. As we want to study the evolution of an initially almost circular particle orbit, we note that we cannot neglect the planetary part of $U_3$, because of the $1/e$ factor in $d\omega/dt$. The two first equation are coupled, Eq. (\ref{dMdt}) will be solved in a second phase after injection of their solution. These equations are nonetheless irregular for small eccentricity regime. Thus, we will use the complex variable $ z = e \exp(i\omega)$ to render them regular \citep{Wyatt}. Moreover, from Eq. (\ref{dedt}), we can deduce that the eccentricity is maximum when $\omega = \omega_p$. This information, combined with the initial value of the Hamiltonian, allows us to compute the maximal eccentricity as a function of $a$. These maximums prove themselves to be less than 0.5 in any case, so that we can linearize the system in $z$ for an easier solving. It yields

\begin{equation}
\frac{dz}{dt} = i\left((A_{B,1}+A_{p,1})z-A_{p,2}\right)\qquad,~ z = e \exp(i\omega),
\end{equation}

where

\begin{align}
A_{B,1} &= \frac{3}{16} n \left(\frac{a_B}{a}\right)^2 \left(\frac{3}{2}e_B^2 + 1\right)\quad;\\
A_{p,1} &= \frac{3}{4} \frac{m_p}{m_B} n \left(\frac{a}{a_p}\right)^3 \left(1-e_p^2\right)^{-\frac{3}{2}}\quad;\\
A_{p,2} &= \frac{15}{16} \frac{m_p}{m_B} n e_p \left(\frac{a}{a_p}\right)^4  \left(1-e_p^2\right)^{-\frac{5}{2}}\quad.
\end{align}

We now solve the system and get the eccentricity, precession and mean anomaly as a function of time. For null initial eccentricity, it is written

\begin{align}
e(t) =&~ \frac{2A_{p,2}}{A_{B,1}+A_{p,1}}\left|\sin\left(\frac{(A_{B,1}+A_{p,1})t}{2}\right)\right|\quad;\\
\omega(t) =&~ \frac{A_{B,1}+A_{p,1}}{2}t \pmod \pi - \frac{\pi}{2} + \omega_p\quad;\\
M(t) =&~ (n + A_{B,1}-\frac{7}{3}A_{p,1}+\frac{1}{8}A_{p,2})t + M(0)\quad.
\end{align}

If we represent the motion of $z$ on the complex plane, we get exactly the circle depicted in Fig. 2 of \cite{Wyatt}. These formula were used to generate Fig. \ref{fig:spirals}.

\section{Density of stars around HD~106906}
\label{AppendixDensity}

The first step to investigate the density of stars around the HD~106906 system is to build a complete list of known members in the LCC subgroup of the Sco-Cen association. Our list of LCC members is based on previous surveys of this region \citep{HipparcosOB,Preibisch2008,Song2012,Pecaut2016} and consists of 369~stars. In the following, we estimate the current density of stars around the planetary system and its evolution in time. Thus, our methodology requires prior knowledge of the distances, proper motions and radial velocities for the individual stars in our sample. 

The Tycho-Gaia Astrometric Solution \citep[TGAS,][]{Lindegren2016} from the Gaia data release 1 provides trigonometric parallaxes and proper motions for 203~stars in our sample. To access more proper motion data, we also searched for this information in the PPMXL \citep{Roeser2010}, UCAC4 \citep{Zacharias2012} and SPM4 \citep{Girard2011} catalogs. Doing so, we find proper motions for 368 stars of the sample. We use the TGAS proper motions for the 203 stars and take the weighted mean of the multiple measurements given by the other catalogs (PPMXL, UCAC4 and SPM4) for the remaining 165~stars. Then, we searched the SIMBAD/CDS databases \citep{Wenger2000} for radial velocity information using the data mining tools available on the site. The radial velocities that we use in this work come from \citet{Wilson1953}, \citet{Duflot1995}, \citet{Brossat2000}, \citet{Torres2006}, \citet{Gontcharov2006}, \citet{Holmberg2007}, \citet{Mermilliod2009}, \citet{ScoCen}, \citet{Song2012}, \citet{Kordopatis2013} and \citet{Desidera2015}. We found radial velocity for 184~stars of our sample.

We apply the methodology developed by \citet{BailerJones2015} to convert parallaxes into distances (see Sect.~7 of his paper). The systematic errors of about 0.3 mas in the TGAS parallaxes reported by \cite{Lindegren2016} were added quadratically to the parallax uncertainties. The three-dimensional position of the stars are calculated from the individual distances in a $XYZ$ grid where $X$ points to the Galactic center, $Y$ points in the direction of Galactic rotation, and $Z$ points to the Galactic north pole. The reference system has its origin at the Sun. Then, we use the procedure described in \citet{Johnson1987} to compute the $UVW$ components of the spatial velocity for each star that are given in the same reference system. We perform a 3$\sigma$ clipping in the distribution of proper motions, parallaxes, radial velocities and spatial velocities to remove obvious outliers. This procedure reduces the dataset to a total of 312~stars, but only 141~stars in this sample exhibit published radial velocities and 102~stars have complete data (proper motions, radial velocities and parallaxes). Based on this subset of 102~stars we calculate a revised mean spatial velocity of the LCC association  of $(U,V,W)=(-8.5,-21.1,-6.3)\pm(0.2,0.2,0.2)$~km/s (not corrected for the solar motion). 

We note that 39~stars in the sample of 141~stars with known radial velocities do not have published parallaxes in the TGAS catalog. Individual parallaxes (and distances) can be inferred for these stars from the moving-cluster method under the assumption that they are co-moving. This method uses proper motions, radial velocities and the convergent point position of the moving group to derive individual parallaxes for group members \citep{Galli2012}. We emphasize that the so-derived kinematic parallaxes are meaningful and provide valuable information in this work to increase the number of stars with measured parallax in our sample. We adopt the space motion listed above and the formalism described in Sect.~2 of \citet{Galli2016} to estimate the convergent point position and the kinematic parallaxes for each group member. Using a velocity dispersion of $\sigma_{v}=1.5$~km/s and distance estimate of 120~pc for the LCC association \citep[see e.g.,][]{deBruijne1999} we find a convergent point solution located at $(\alpha_{cp},\delta_{cp})=(104.8^{\circ},-37.2^{\circ})\pm(1.0^{\circ},0.8^{\circ})$ with chi-squared statistics $\chi^{2}_{red}=0.92$ and correlation coefficient of $\rho=-0.98$. To gain confidence in the so-derived kinematic parallaxes we compare our results with the trigonometric parallaxes from the TGAS catalog for the stars in common. We find a mean difference of $0.1$~mas and r.m.s. of $0.6$~mas, that are significantly smaller than the typical error on the kinematic parallaxes ($\sim 0.8$~mas) derived from the moving-cluster method in this analysis. This confirms the good agreement between the two datasets. Thus, the final sample with complete information (proper motion, radial velocity and parallax) that we use in this work to estimate the early density of stars around HD~106906 consists of 141~stars.  

\begin{figure}[!h]
\begin{center}
\includegraphics[width=0.8\linewidth]{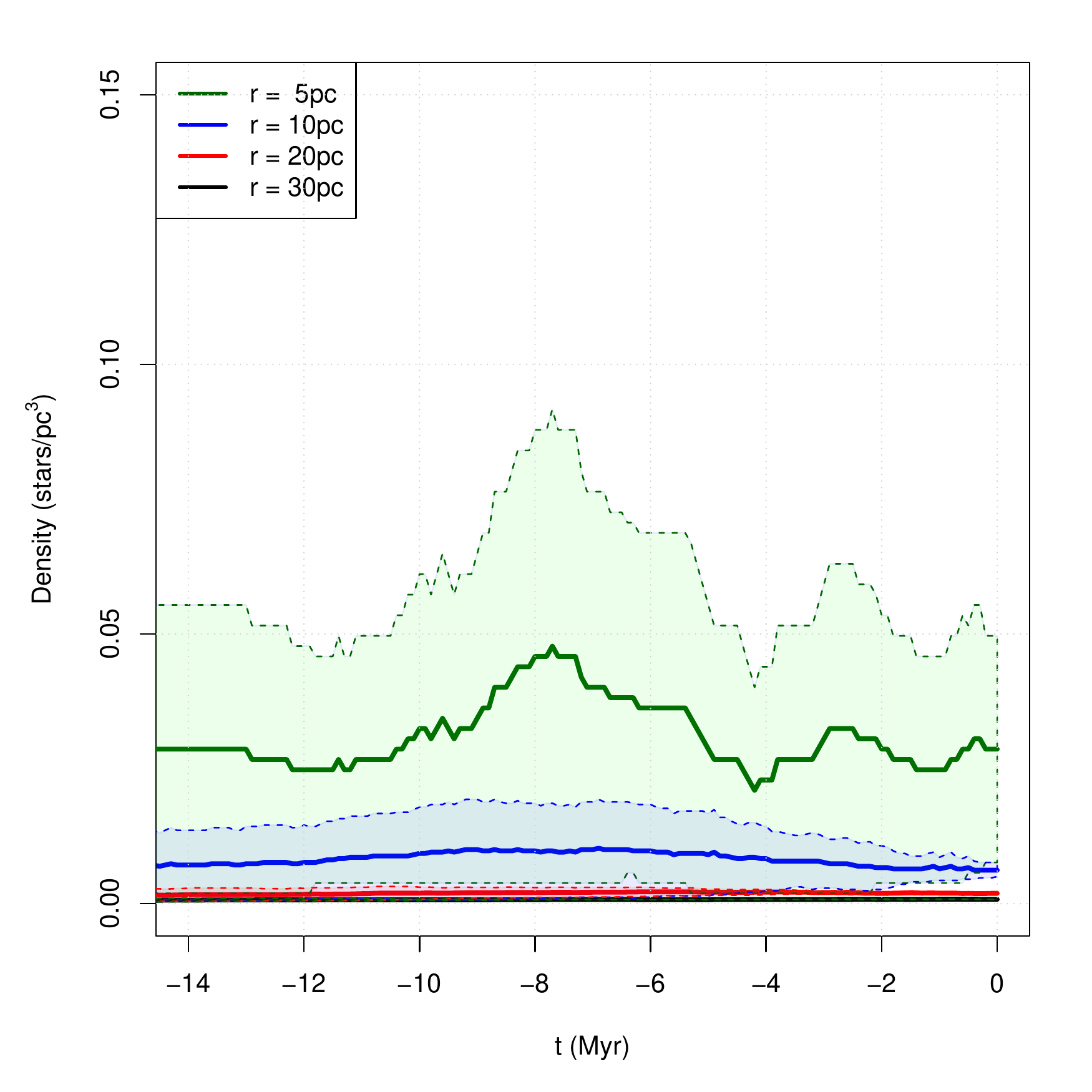}
\caption{
\label{figC1}
Evolution of the density of stars for different radii around the HD~106906 planetary system. The colored regions indicate the upper and lower limits (at the $1\sigma$ level) for the density of stars at a given radius. }
\end{center}
\end{figure}

In a subsequent analysis, we consider the present day location of the 141~stars and use the $UVW$ spatial velocity for each star to calculate their $XYZ$ positions backward in time. We compute the stellar positions as a function of time in steps of $0.1$~Myr from $t=0$ (current position) to $t=-14.0$~Myr. The latter value is chosen to be consistent with an upper limit for the age estimate of the HD~106906 system as derived by \citet{LCC} from different evolutionary models. Then, we count the number of stars in the vicinity of HD~106906 for different radii ($r=5,10,15,...,30$~pc) and determine the density of stars around the target. Figure~\ref{figC1} illustrates the results of this investigation. Our analysis indicates that the early density of stars around HD~106906  (at $t=-7.7$~Myr) was higher than the current value by a factor of about 1.7 for $r=5$~pc. At this stage it is important to mention that our result for density of stars is restricted to known members of the LCC association with complete data in our sample for which we can calculate spatial velocities and compute their positions back in time. As soon as new data (parallaxes and radial velocities) from the upcoming surveys (e.g., Gaia) become available and other group members are identified, a more refined analysis of this scenario will be made possible.

\begin{figure}[!h]
\begin{center}
\includegraphics[width=0.8\linewidth]{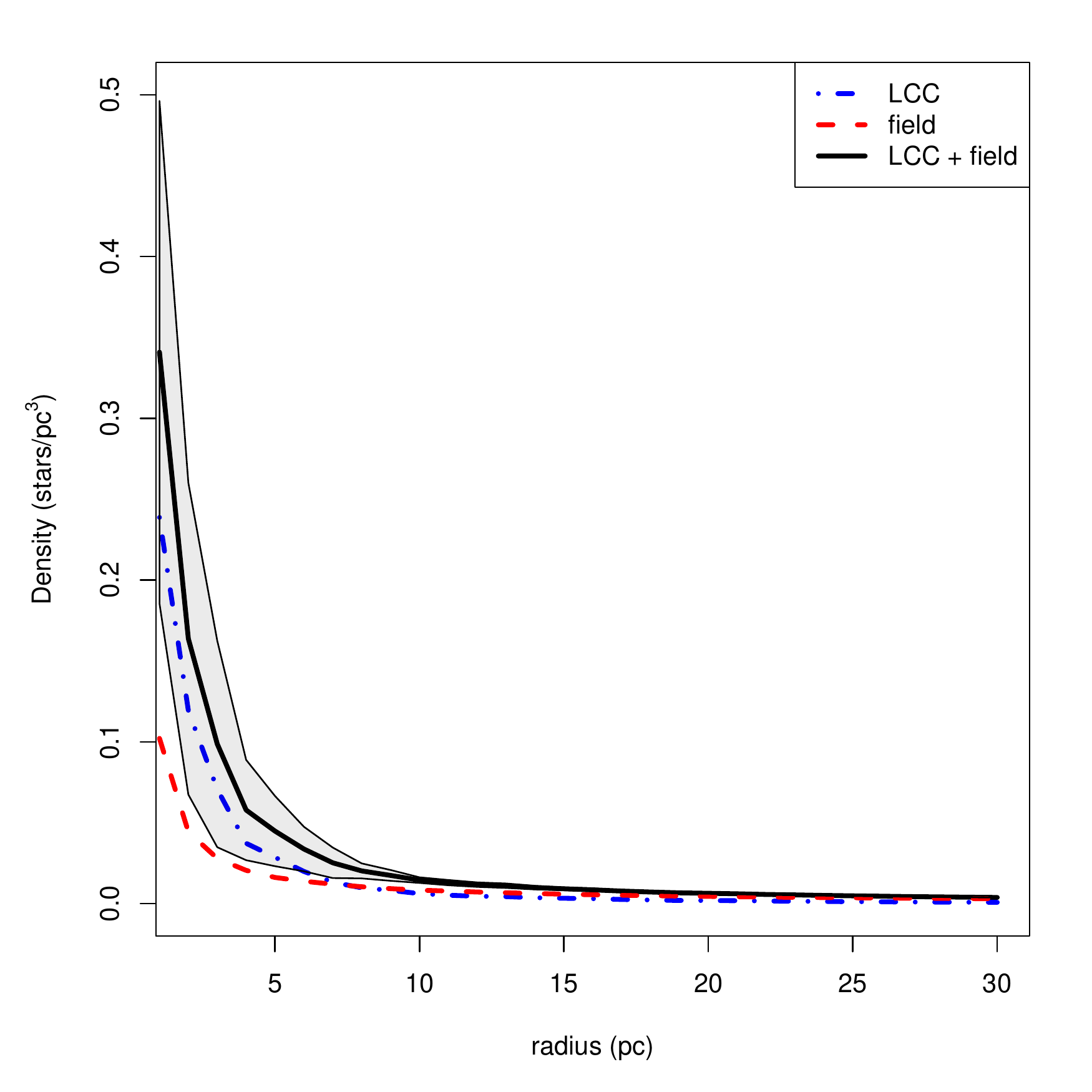}
\caption{
\label{figC2}
Current density of stars in the vicinity of the HD~106906 system inferred from LCC cluster members and field stars. The colored region indicates the upper/lower limits (at the $1\sigma$ level) for the final density of stars (cluster $+$ field) at different radii around the target.  }
\end{center}
\end{figure}

One alternative approach to better constrain the density of stars around HD~106906 consists of investigating the contribution of field stars (not related to the LCC association) in our solution. In this context, we use the model of stellar population synthesis from \citet{Robin2003} to simulate a catalog of pseudo-stars and their intrinsic properties (e.g., distances, spectral types, ages, magnitudes, etc) in the direction of the HD~106906 system. We run the model with a distance range from 0 to 300~pc and a solid angle of 20~deg$^{2}$ centered around the target. These values are chosen to include known members of the LCC association that is clearly spread in angular extent and exhibits significant depth effects along the line of sight. We do not constrain our simulations in magnitudes and spectral types to get a more complete picture of the stellar population in this region. We use a distance step of 0.5~pc in our simulations that is the minimum value that can be used in the model. The synthetic stars are all supposed to be at the same coordinates. So, we run a number of 1000 simulations to generate random coordinates for the simulated stars and use them (together with the distances provided by the model) to calculate the stellar three-dimensional positions in the $XYZ$ grid. Figure~\ref{figC2} shows the density of stars around HD106906 for different radii obtained from our sample of LCC stars, the pseudo-stars from our simulations and a combined result that includes both (cluster + field). Although this analysis cannot be extrapolated backward in time (as in Fig.~\ref{figC1}), it yields a more refined value for the current ($t=0$) density of stars. However, we emphasize that the results obtained for small radii around the target (i.e., $r\leq5$~pc) are calculated with a small number of stars (typically, less than ten~stars) and they should be regarded with caution. Thus, we conclude that the present-day density of stars in the vicinity of the HD~106906 system for $r>5$~pc is $\leq0.07$~stars/pc$^{3}$ (within the $1\sigma$ error bars). We infer from the results presented in Fig.~\ref{figC1} an upper limit of $\sim0.11$~stars/pc$^{3}$ for the density of stars around HD~106906, a result that will need further confirmation as soon as more data becomes available. 

\end{appendix}

\end{document}